\documentclass[runningheads]{llncs}

\usepackage[mobile]{eccv}

\usepackage{eccvabbrv}

\usepackage{graphicx}
\usepackage{booktabs,array}
\usepackage{tabularx}
\usepackage{multirow}
\usepackage{subcaption}
\usepackage{amssymb}
\usepackage{makecell}
\usepackage{mathabx}
\usepackage{wrapfig}
\usepackage[table]{xcolor}

\usepackage[accsupp]{axessibility}  

\newcommand{\topprime}[1]{%
  \mathchoice
    {#1^{\raisebox{-0.60ex}{$\displaystyle\prime$}}}
    {#1^{\raisebox{-0.50ex}{$\textstyle\prime$}}}
    {#1^{\raisebox{-0.35ex}{$\scriptstyle\prime$}}}
    {#1^{\raisebox{-0.25ex}{$\scriptscriptstyle\prime$}}}
}

\definecolor{fst}{rgb}{0.909, 0.504, 0.588}
\definecolor{sec}{rgb}{0.994, 0.806, 0.742}
\definecolor{thd}{rgb}{0.999, 0.966, 0.921}
\definecolor{lin}{rgb}{0.32, 0.27, 0.40}

\usepackage{hyperref}

\usepackage{orcidlink}

\begin{document}

\title{Denoising the Deep Sky: Physics-Based CCD Noise Formation for Astronomical Imaging} 

\titlerunning{Denoising the Deep Sky}

\author{
Shuhong Liu\inst{1,2} \and
Xining Ge\inst{2} \and
Ziying Gu\inst{1} \and
Quanfeng Xu\inst{3} \and
Lin Gu\inst{4} \\
Ziteng Cui\inst{1,2} \and
Xuangeng Chu\inst{1,2} \and
Jun Liu\inst{2} \and
Dong Li\inst{2} \\
Tatsuya Harada\inst{1,5}
}

\authorrunning{Liu et al.}

\institute{$^1$The University of Tokyo, Tokyo, Japan\\
$^2$I2WM, Tokyo, Japan\\
$^3$Shanghai Astronomical Observatory, Shanghai, China\\ $^4$Tohoku University, Sendai, Japan\\
$^5$RIKEN AIP, Tokyo, Japan}

\maketitle

\begin{abstract}
Astronomical imaging remains noise-limited under practical observing conditions. Standard calibration pipelines remove structured artifacts but largely leave stochastic noise unresolved. Although learning-based denoising has shown strong potential, progress is constrained by scarce paired training data and the requirement for physically interpretable models in scientific workflows. We propose a physics-based noise synthesis framework tailored to CCD noise formation in the telescope. The pipeline models photon shot noise, photo-response non-uniformity, dark-current noise, readout effects, and localized outliers arising from cosmic-ray hits and hot pixels. To obtain low-noise inputs for synthesis, we stack multiple unregistered exposures to produce high-SNR bases. Realistic noisy counterparts synthesized from these bases using our noise model enable the construction of abundant paired datasets for supervised learning. Extensive experiments on our real-world multi-band dataset curated from two ground-based telescopes demonstrate the effectiveness of our framework in both photometric and scientific accuracy. Code is available at \url{https://github.com/ShuhongLL/Denoising-Deep-Sky}.
\end{abstract}

\section{Introduction}
\label{sec:intro}

Modern imaging systems devote substantial effort to suppressing sensor noise through advances in both hardware design and computational methods. Nevertheless, digital images remain affected by fluctuations intrinsic to photon detection, electronic readout, and residual artifacts introduced by sensor architecture \cite{pratt2007ch5}. Reducing these noise sources while faithfully preserving signal fidelity has therefore become a central objective for computational imaging.

For consumer photography, particularly in extreme low-light conditions, learning based denoising has advanced substantially \cite{chen2018learning}. To mitigate the scarcity in acquiring strictly paired data, recent research has focused on developing sophisticated noise modeling and synthesis techniques \cite{zhang2021rethinking,wei2021physics,monakhova2022dancing,jin2023lighting,feng2023learnability,jiang2025revolutionizing}. Within this setting, physics-based models \cite{wei2020physics,cao2023physics,li2025noise}, implicit modeling \cite{feng2023physics,zhang2023towards}, and non-parametric approaches \cite{colom2014nonparametric,mosleh2024non} have been proposed to synthesize camera-dependent noisy RAW samples at scale, tailored to the ISO-aware noise characteristics of the modern Complementary
Metal-Oxide-Semiconductor (CMOS) sensors that are prevalent in consumer cameras.

\begin{figure}[!tp]
    \begin{center}
        \includegraphics[width=\textwidth]{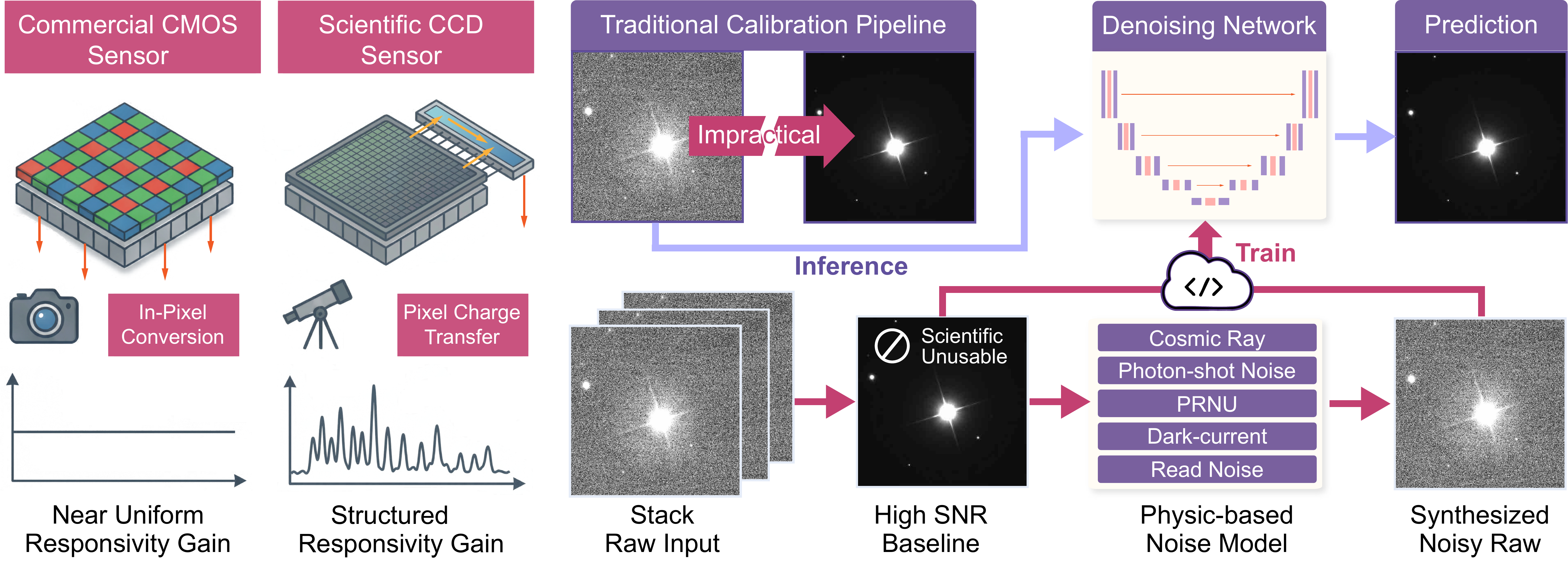}
    \end{center}
    \caption{Overview of denoising pipeline. Left: a CMOS sensor performs in-pixel conversion and shows a near-uniform pixel responsivity gain, while a scientific CCD sensor transfers pixel charge and exhibits spatially structured gain. Right: paired training images are limited since clean signals cannot be recovered by extending exposure or traditional calibration. We stack raw inputs to form a high-SNR baseline, synthesize realistic noisy raws using a physics-based noise model, and train the denoising network.}
    \label{fig1:overview}
\end{figure}

Scientific imaging, particularly in astronomy, presents a different landscape where Charge-Coupled Device (CCD) sensors remain the dominant technology \cite{howell2006handbook}. The CCD architecture differs fundamentally from CMOS cameras. Instead of embedding amplifiers and converters at the pixel level, CCDs transfer charge serially to a small set of output amplifiers \cite{janesick1987scientific}. This readout process introduces distinctive noise signatures \cite{faraji2006ccd}, including segment-specific bias and gain variation \cite{lauer1989reduction}, charge transfer inefficiency \cite{massey2010pixel}, clock-induced charge \cite{janesick2007photon}, and bias drift and striping \cite{grogin2011acs}. At longer wavelengths, back-illuminated CCDs exhibit interference fringes or etaloning \cite{howell2006handbook}, caused by internal reflections within the silicon substrate. Unlike consumer CMOS sensors, science CCDs record in a single spectral band defined by filters rather than through a mosaic color pattern, which alters the statistical structure of the data. Astronomical CCDs are also exposed to unique observational effects, including cosmic-ray strikes \cite{van2001cosmic} and persistent hot pixels \cite{mclean2008electronic} that introduce localized outliers.


Unlike consumer CMOS imaging, where noise can often be reduced by simply extending the exposure time, astronomical observations are fundamentally constrained by atmospheric effects and target dynamics. Ground-based telescopes are subject to turbulence-induced wavefront aberrations \cite{fried1966optical}, scintillation \cite{roddier1981v}, intrinsic source variability \cite{zhai2014detection}, and Earth’s rotation \cite{benn1998brightness}, all of which evolve on timescales that make long integrations for obtaining high-SNR images impractical. As a result, CCD observations are typically acquired as a sequence of shorter exposures that are individually calibrated and then registered and stacked \cite{fruchter2002drizzle}. While stacking improves SNR and enables the rejection of outliers, it also faces notable limitations. Residual offsets can arise from imperfect telescope tracking and co-registration. \cite{bertin2010swarp}. Moreover, the strict alignment required for stacking is time-consuming and not always feasible \cite{stoughton2002sloan}, especially for faint or time-variable sources \cite{alard1998method}, or for large-area survey observations where rapid coverage is prioritized \cite{ivezic2019lsst}. Consequently, the scientific usability of individual exposures remains essential, while conventional calibration pipelines primarily suppress structured artifacts and leave stochastic noise largely unresolved \cite{howell2006handbook}.

To learn the denoising model for astronomical imaging and overcome the data sparsity commonly seen in scientific imaging, we propose a physics-based noise synthesis framework tailored to CCD noise formation. As illustrated in \Cref{fig1:overview}, our pipeline models the principal sources of CCD noise, including photon shot noise, per-amplifier bias offsets and gain non-uniformity, dark-current noise, and readout effects such as read noise and charge transfer inefficiency at the output amplifier. It further accounts for observational phenomena, including transient cosmic-ray hits and persistent hot pixels, which can noticeably affect raw exposures. A natural question is how to obtain the ``clean'' bases for synthesis, since obtaining truly noise-free references is impractical \cite{guo2026deeper}. We leverage common practice in modern telescopes that track targets with active guiding so the scene remains approximately fixed on the detector across sequential exposures. We directly stack multiple unregistered observation frames, which inevitably broadens the astrophysical point-spread function due to residual drift and seeing variation, but substantially suppresses stochastic noise without introducing interpolation artifacts. Such averaged images are not scientifically usable per se; however, they provide abundant high-SNR bases that preserve correct large-scale background structure and instrument signatures. Our synthesis pipeline can generate realistic noisy counterparts from these bases, enabling the construction of paired datasets for supervised learning. To validate the framework, we further construct a real astronomical image dataset that contains paired raw noisy frames and instrument-pipeline calibrated frames collected from two twin ground-based telescope observations for real-world evaluation.

Our main contributions can be summarized as follows:
\begin{itemize}
    \item We propose a physics-based noise formulation model tailored to astronomical CCD imaging, capable of synthesizing realistic noisy RAW data that closely matches the statistical properties of real observations.  
    \item We construct real-world multi-band astronomical imaging datasets from ground telescope observations, providing both raw and processed observations for noise synthesis and quantitative evaluation.  
    \item Comprehensive in-domain and cross-instrument experiments demonstrate that our method consistently outperforms baseline approaches in both photometric fidelity and scientific accuracy.
\end{itemize}

\section{Related Work}
\label{sec:related_work}

\subsection{Denoising in Consumer CMOS Imaging}

Recent efforts in CMOS image denoising have largely shifted from traditional model-based methods to deep learning approaches. While early learning-based methods trained on synthetic Gaussian noise showed promise \cite{zhang2017beyond}, they often struggled to generalize to real-world scenarios where noise is more complex and signal-dependent \cite{chen2018learning,liu2026rawild}. To bridge this gap, physically based noise models that account for signal-dependent and independent noise have been proposed \cite{wei2020physics,jin2023lighting,zhang2021rethinking,feng2023learnability}. Further refinements to these models have incorporated ISO-correlated noise characteristics \cite{cao2023physics} and explored nonparametric approaches \cite{mosleh2024non} to capture intricate noise distributions. To address the challenge of acquiring large-scale paired datasets, unsupervised and self-supervised approaches have been developed without requiring clean images \cite{lehtinen2018noise,zou2023iterative,pan2023random}. Beyond single-image methods, burst-based methods \cite{mildenhall2018burst,godard2018deep} exploit temporal redundancy across multiple exposures to further improve denoising quality. More recently, generative priors have been explored for zero-shot denoising without training \cite{qin2024el2nm,shi2024zero}.

\subsection{Denoising in Astronomical CCD Imaging}

 Astronomical imaging denoising is a fundamental challenge due to the extremely low photon counts associated with observing faint celestial objects. A cornerstone of conventional practice is the use of calibration frames, which correct for readout patterns, thermal signals, pixel-to-pixel nonuniformity, and optical vignetting \cite{janesick1987scientific,howell2006handbook}. To mitigate transient artifacts and improve the SNR, multiple dithered exposures, sigma-clipping, and median filtering are commonly employed \cite{van2001cosmic,van2012cosmic}, while advanced techniques such as drizzle have been developed to address undersampled images from the Hubble Space Telescope \cite{fruchter2002drizzle}. In addition, transform-domain methods, especially wavelet-based approaches, separate noise from astronomical structures by exploiting multi-scale representations \cite{starck2007astronomical}. More recently, deep learning has emerged as a promising approach, often bypassing the need for a multi-stage calibration pipeline by performing end-to-end denoising that simultaneously removes instrumental signatures and artifacts \cite{zhang2020deepcr,vojtekova2021learning,liu2026fluxflow}. Self-supervised methods are also proposed to model noise directly from observations, offering an automated solution for the modern sky survey \cite{zhang2022noise2astro,liu2025astronomical,guo2026deeper}.

\begin{figure}[!tp]
    \begin{center}
        \includegraphics[width=\textwidth]{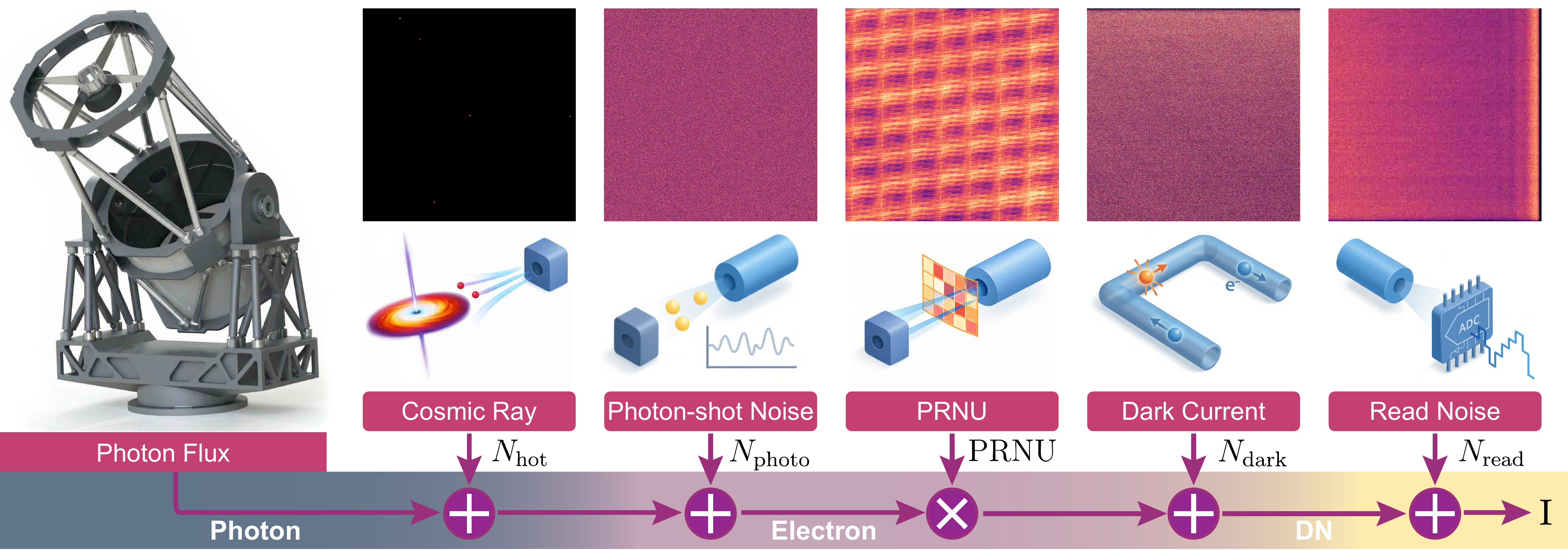}
    \end{center}
    \caption{Overview of noise formation in astronomical imaging. Unique telescope optics and CCD science cameras introduce noise characteristics that are distinct from commercial CMOS imaging. Zoom in the noise frames for better visibility.}
    \label{fig2:noise_formation}
\end{figure}

\section{Method}
\label{sec:method}
To construct pair-wise training data for learning denoising models, our objective is to synthesize realistic CCD noise using only a limited number of stacked clean bases and calibration frames. We begin by introducing the CCD image formation process in \Cref{sec:ccd_image_formation}, then describe the synthesis of signal-dependent noise and signal-independent noise in \Cref{sec:signal_dependent_noise} and \Cref{sec:signal_independent_noise}. The overview of the noise formation model is illustrated in \Cref{fig2:noise_formation}.


\subsection{Astronomical Image Formation Model}
\label{sec:ccd_image_formation}
A CCD camera converts incident photons into photoelectrons and also accumulates thermally generated dark current during the exposure. The total charge collected in each pixel is then transferred through the serial register to one or more output amplifiers and digitized. At the frame level, this process can be separated into signal-dependent terms, which scale with illumination and exposure time, and signal-independent terms, which originate from the readout chain, sparse defects, and quantization. The system gain is $G$ in electrons per Analog-to-Digital Unit (ADU). Let $S(x)$ denote the photon arrival rate at pixel $x$ in photons per second, $\eta(x)$ the quantum efficiency in electrons per photon, and $\mathrm{PRNU}(x)$ a multiplicative factor describing pixel-to-pixel Photo Response Non-uniformity (PRNU). The raw digital value $I(x)$ is then modeled as follows:
\begin{align}
\label{eq:noise_formation}
I(x) &= 
\underbrace{N_{\rm overscan}(x)}_{\text{Constant}} \nonumber + \underbrace{\frac{1}{G}\,\bigl(
t \eta(x) S(x) \mathrm{PRNU}(x) + N_{\rm photon}(x) + N_{\rm dark}(x)\bigr)}_{\text{Signal-dependent}} \nonumber \\
&+ \underbrace{N_{\mathrm{read}}(x) + N_{\rm digit}(x) + N_{\mathrm{hot}}(x)}_{\text{Signal-independent}}
\end{align}
In this expression, $N_{\rm overscan}(x)$ is the overscan pedestal, which provides a constant offset ensuring that all digitized pixel values remain positive. The signal-dependent component consists of three contributions, including the target signal describing the expected photoelectron count, the photon-shot noise term $N_{\rm photon}(x)$, and the dark-current noise realization $N_{\rm dark}(x)$. The factor $1/G$ provides the linear conversion from electrons to the ADU domain. The signal-independent component then comprises additive $N_{\rm read}(x)$, $N_{\rm hot}(x)$, and $N_{\rm digit}(x)$. Here, $N_{\rm read}(x)$ is read noise arising from the output amplifier and correlated-double-sampling chain, including contributions from reset and thermal noise and low-frequency amplifier fluctuations. $N_{\rm digit}(x)$ is the digitization error. $N_{\rm hot}(x)$ is a sparse defect term that aggregates hot-pixel outliers and cosmic-ray events.

\begin{figure}[!tbp]
\centering
\setlength{\tabcolsep}{0pt}
\begin{tabular}{@{} m{0.04\columnwidth} *{4}{m{0.198\columnwidth}} @{}}
  & \multicolumn{4}{@{}l@{}}{%
    \includegraphics[width=0.5\linewidth]{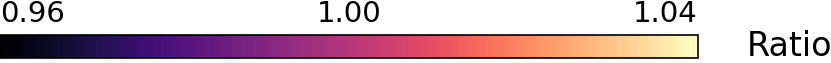}%
  } \\
  \centering\rotatebox{90}{\scriptsize $\topprime{G}$ band} &
  \includegraphics[width=0.96\linewidth]{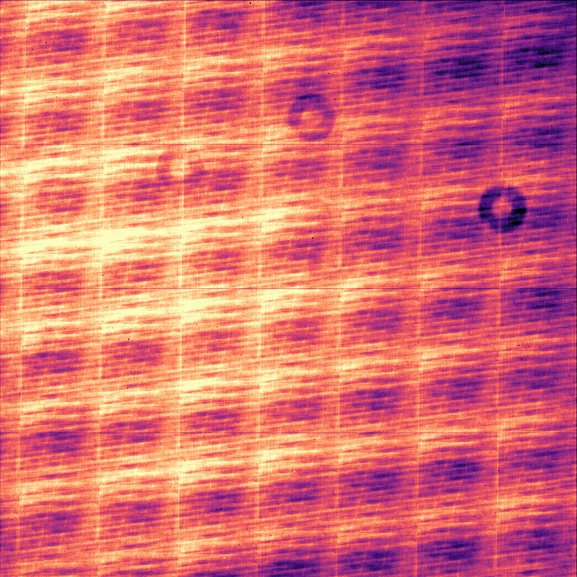} &
  \includegraphics[width=0.96\linewidth]{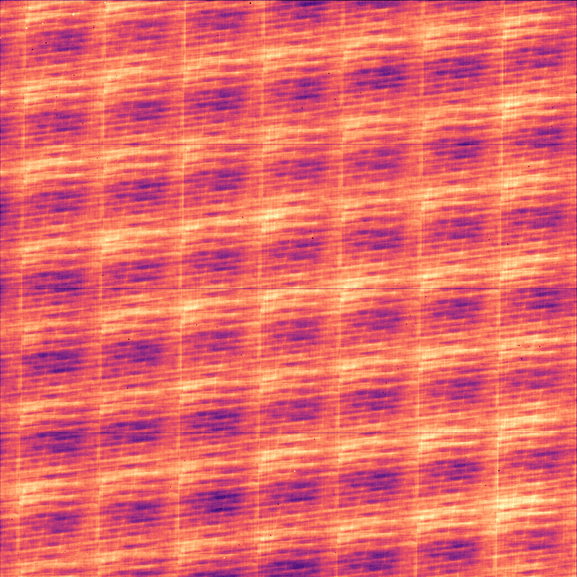} &
  \includegraphics[width=0.96\linewidth]{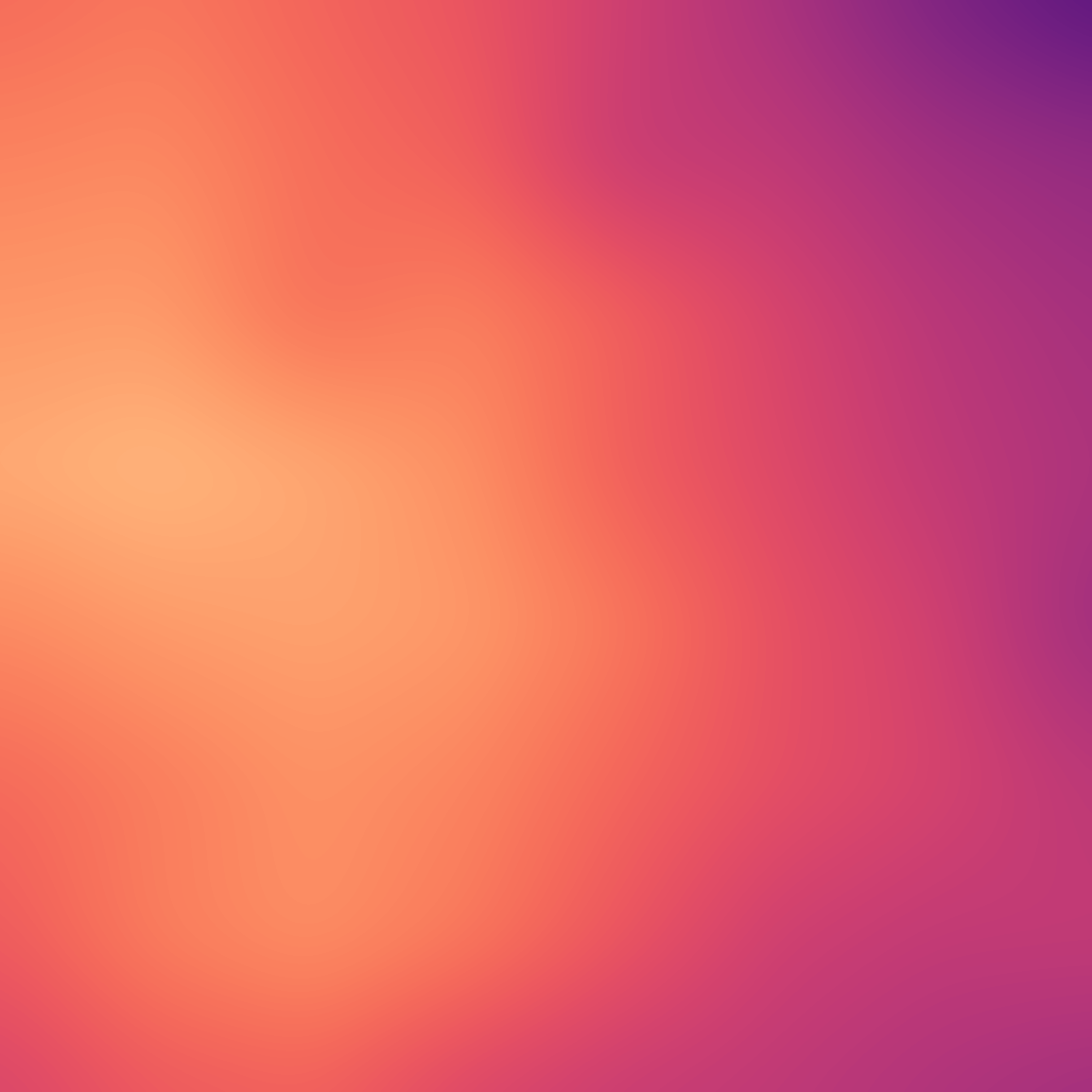} &
  \includegraphics[width=0.96\linewidth]{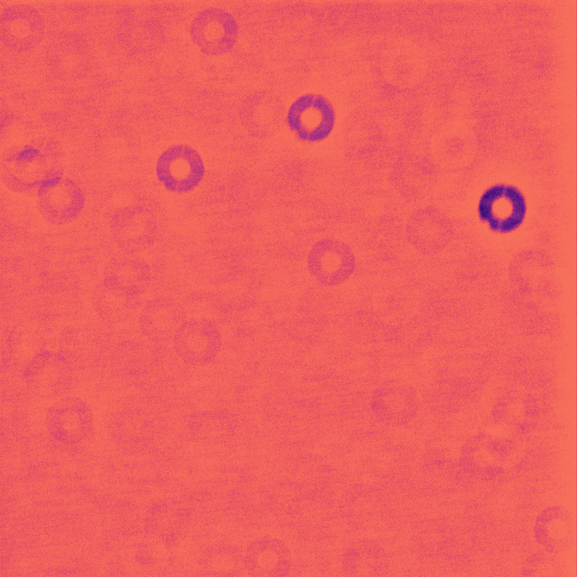} \\

  \centering\rotatebox{90}{\scriptsize $\topprime{R}$ band} &
  \subcaptionbox{Skyflat\label{fig3:prnu:a}}[0.19\columnwidth]{%
    \includegraphics[width=\linewidth]{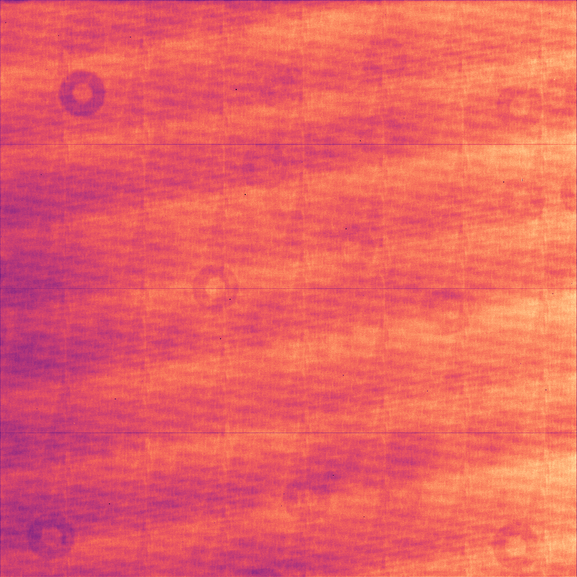}} &
  \subcaptionbox{Fixed Pattern\label{fig3:prnu:b}}[0.19\columnwidth]{%
    \includegraphics[width=\linewidth]{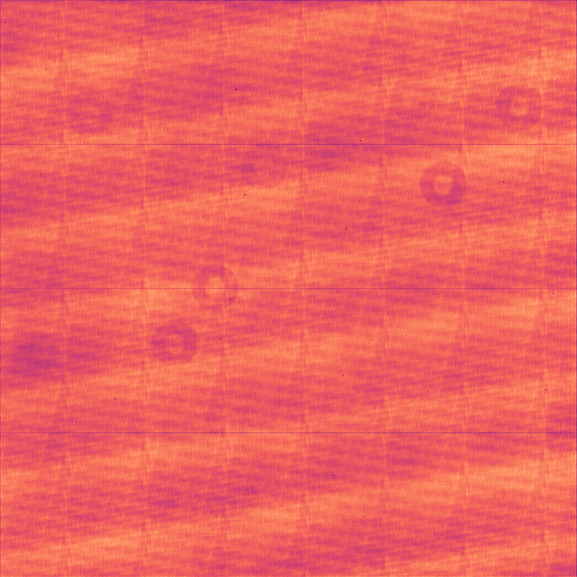}} &
  \subcaptionbox{Twilight\label{fig3:prnu:c}}[0.19\columnwidth]{%
    \includegraphics[width=\linewidth]{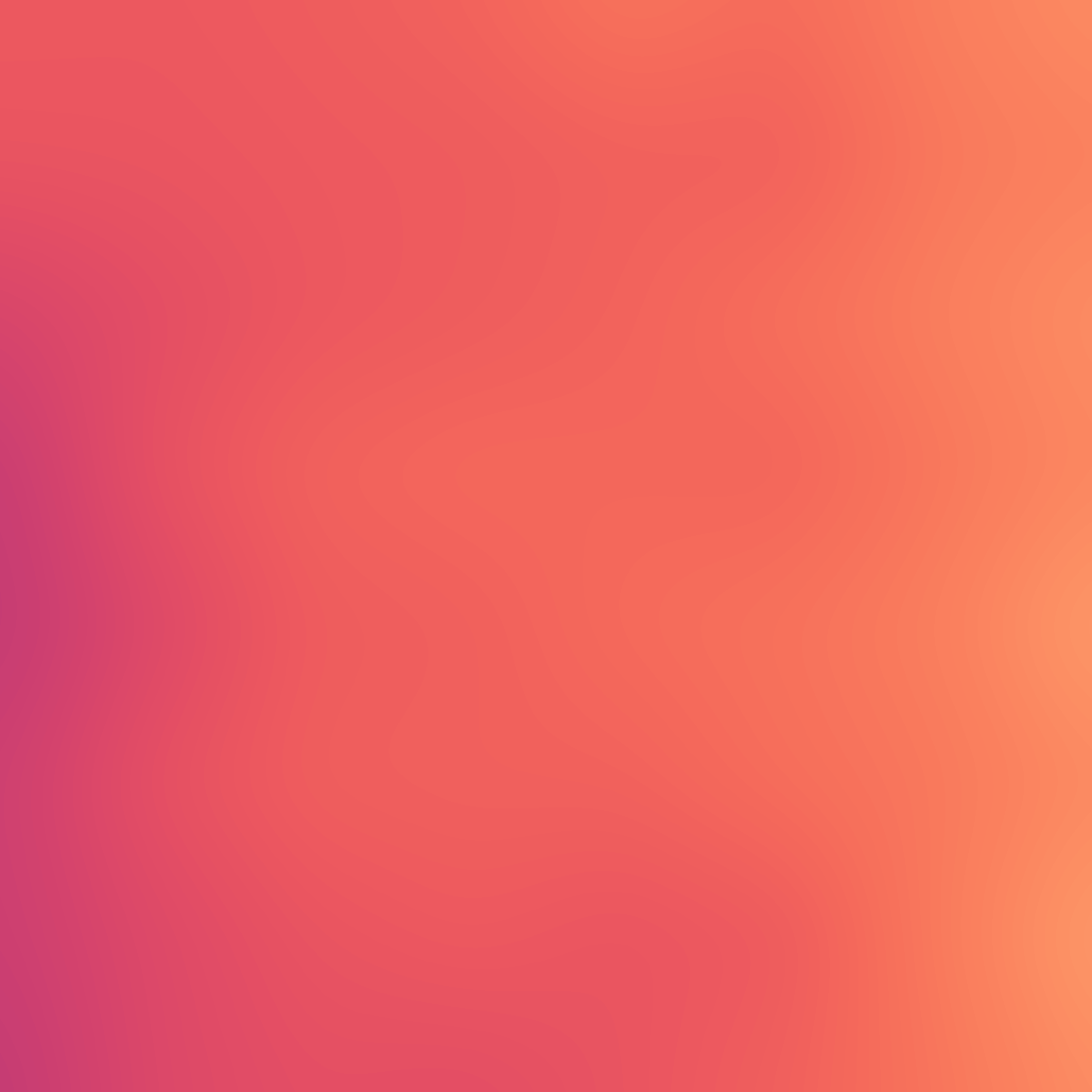}} &
  \subcaptionbox{Donuts\label{fig3:prnu:d}}[0.19\columnwidth]{%
    \includegraphics[width=\linewidth]{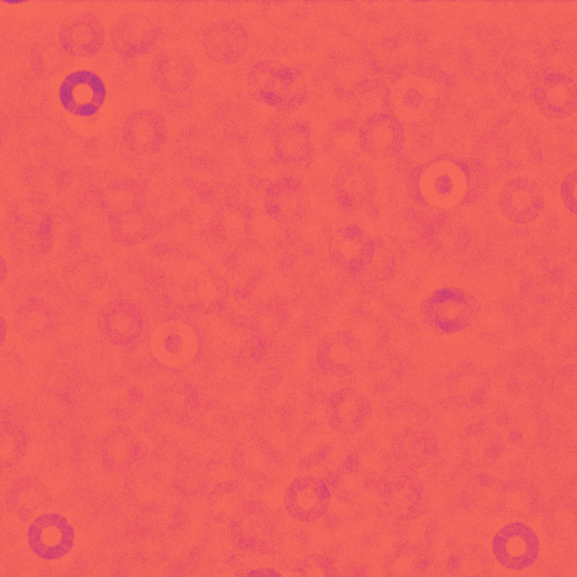}}
\end{tabular}
\caption{PRNU comparison between $\topprime{G}$ and $\topprime{R}$ bands. A captured sky flat frame (a) is typically expressed as the product of (b) the anti-fringing fixed pattern, (c) the low-frequency twilight sky background, and (d) dust-donut artifacts. PRNU is given by (b) $\times$ (d), where (c) is the external signal.}
\label{fig3:prnu}
\end{figure}

\subsection{Signal-Dependent Noise}
\label{sec:signal_dependent_noise}

\subsubsection{Photo Response Non-Uniformity (PRNU)}
PRNU describes fixed, spatially varying differences in pixel responsivity, such that identical uniform illumination can yield different pixel outputs \cite{howell2006handbook}. In commercial CMOS cameras, this effect is handled through \emph{flat-field correction} or \emph{shading correction}, which uses a uniform bright-field to compensate pixel-response variations \cite{seibert1998flat,basler2024ffc}. In consumer imaging pipelines, PRNU rarely dominates visual quality and is often neglected \cite{wei2021physics,becker2022uniformity,li2025noise}, as in-camera ISP steps typically mask subtle nonuniformity.

In astronomical CCD imaging, PRNU correction is essential for reliable photometry and background estimation: science targets are often faint and extended, and percent-level sensitivity variations can translate into systematic flux biases and large-scale background artifacts \cite{howell2006handbook,massey1997iraf,astropy2024guide}. The observed PRNU pattern reflects both (i) intrinsic detector nonuniformity, such as pixel-to-pixel gain and quantum-efficiency variation, and (ii) multiplicative effects introduced by the optical path. For (i), thinned and back-illuminated CCDs can exhibit wavelength-dependent optical etaloning \cite{walsh2003modelling}. Fringe-suppression manufacturing processes therefore introduce controlled thickness variation to disrupt the Fabry--P\'erot cavity \cite{born2013principles} and reduce fringing, while leaving residual high-frequency structure in the flat-field response \cite{howell2006handbook,howell2012fringe} as shown in \Cref{fig3:prnu:b}. For (ii), dust on the dewar window, filters, or nearby optics introduces out-of-focus shadow features that appear as ring-like \textit{dust donuts}, shown in \Cref{fig3:prnu:d}, in flat fields \cite{howell2006handbook,stsci2024flats,astropy2024guide}.

To correct these multiplicative response variations in real observations, flat-field nonuniformity is typically performed using twilight \emph{skyflats} in the traditional calibration pipeline \cite{mccully2018real}. The telescope is pointed to a blank sky region and multiple exposures are acquired under near uniform illumination. Each skyflat frame is normalized by its median to remove exposure-to-exposure brightness differences, then combined to form a master skyflat frame. A science image is typically calibrated by dividing the raw measurement by the master skyflat map.

However, twilight skyflats contain both the intrinsic detector response and external illumination structure, so direct stacking can bias PRNU estimation. Large-scale twilight gradients, vignetting, and scattered light vary across frames, whereas $\mathrm{PRNU}(x)$ is a fixed sensor property. To isolate the pixel-scale fixed pattern, we remove the low-frequency illumination component from each skyflat frame. Given a skyflat image $N_{\rm sky}^i(x)$, we compute the median-normalized frame $\tilde{N}_{\rm sky}^i(x)$, estimate an illumination background $N_{\rm bg}^i(x)$ from block-wise large kernel median pooling, and form a high-pass residual $H_{\rm hp}^i(x)$ as:
\begin{equation}
H_{\rm hp}^i(x)=\frac{\tilde{N}_{\rm sky}^i(x)}{N_{\rm bg}^i(x)},\qquad
\mathrm{PRNU}_{\rm fp}(x)=\mathrm{norm}\!\left(\mathrm{median}_{\sigma\text{-clip}}\Big(\big\{H_{\rm hp}^i(x)\big\}_i\Big)\right).
\end{equation}
Here $\mathrm{median}_{\sigma\text{-clip}}(\cdot)$ denotes pixel-wise sigma-clipped median aggregation over multiple residuals, and $\mathrm{norm}(\cdot)$ rescales the resulting map to unit median. The resulting $\mathrm{PRNU}_{\rm fp}(x)$ models the static sensor-intrinsic fixed pattern.

After estimating $\mathrm{PRNU}_{\rm fp}(x)$, we extract structured flat-field donut artifacts that remain after fixed-pattern removal. For each normalized skyflat $\tilde{N}_{\rm sky}^i(x)$, we divide out $\mathrm{PRNU}_{\rm fp}(x)$ to obtain a donut map $\mathrm{PRNU}_{\rm donut}^i(x)$, which is randomly sampled during synthesis. The final PRNU model is defined as:
\begin{equation}
\mathrm{PRNU}_{\rm donut}^i(x)=\frac{\tilde{N}_{\rm sky}^i(x)}{\mathrm{PRNU}_{\rm fp}(x)}, \quad
\mathrm{PRNU}(x)=\mathrm{PRNU}_{\rm fp}(x) \times \mathrm{PRNU}_{\rm donut}(x).
\end{equation}

\subsubsection{Photon-Shot Noise}
Unlike CMOS camera, where the effective gain varies with ISO and typically requires careful calibration \cite{wei2020physics} or hypothesized gain estimation \cite{li2025noise} for each setting, the system gain $G$ of an astronomical CCD is precisely characterized and known in advance. Under stationary illumination within the linear response regime, the photon-shot electron count $E$ at pixel $x$ over exposure time $t$ is modeled as a Poisson random variable whose mean equals the expected photoelectron yield after quantum conversion and pixel sensitivity:
\begin{equation}
    E(x) + N_{\rm photon}(x) \sim \mathrm{Poisson}(t \eta(x) S(x) \mathrm{PRNU}(x))
\end{equation}

\begin{figure}[!tp]
    \centering
    \hspace*{\fill}
    \begin{subfigure}[t]{0.32\columnwidth}
        \centering
        \captionsetup{margin={0em,0pt}}
        \caption{Fitting}
        \includegraphics[width=\linewidth]{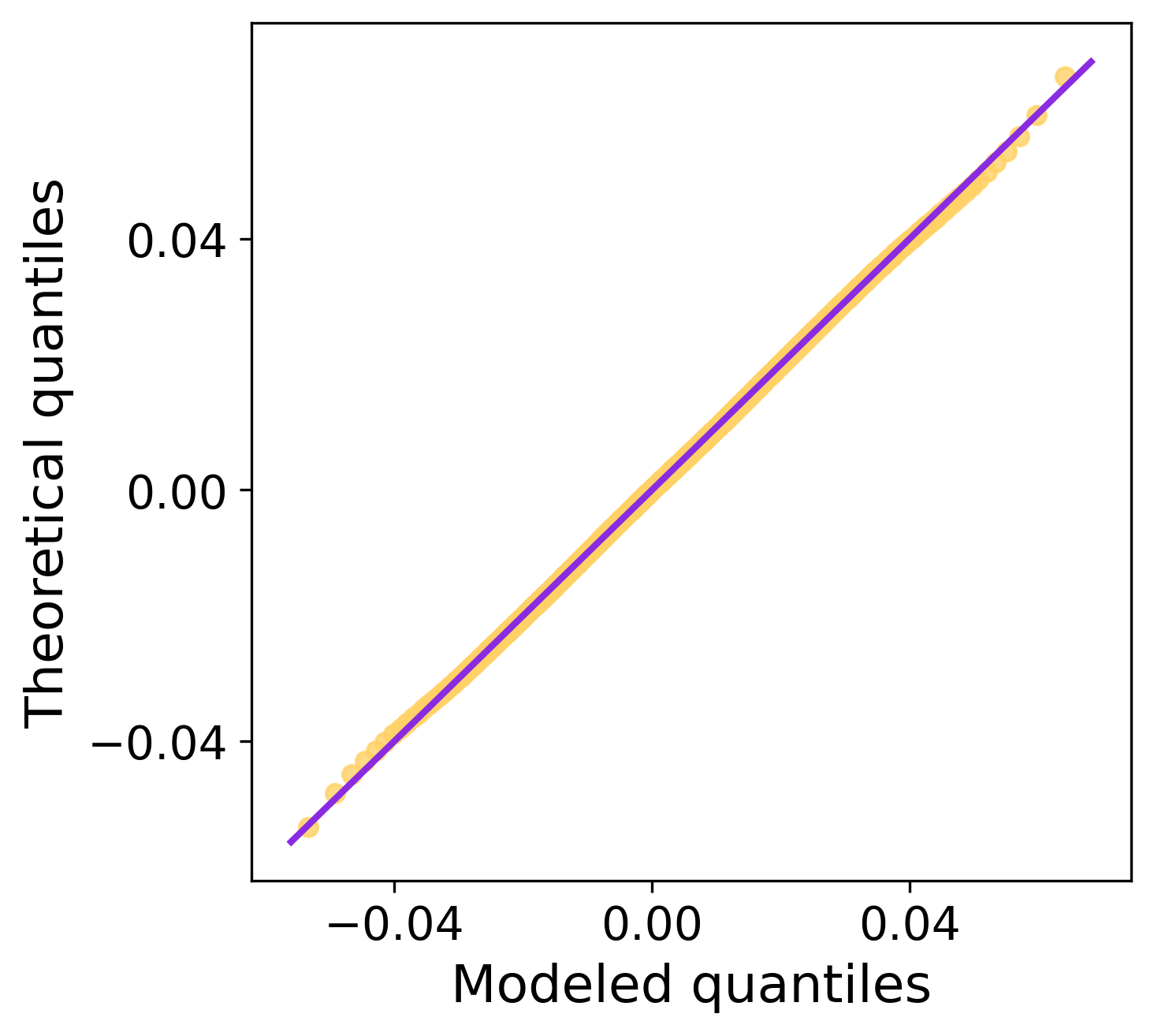}
    \end{subfigure}
    \hspace{0.04\columnwidth}
    \begin{subfigure}[t]{0.32\columnwidth}
        \centering
        \captionsetup{margin={0em,0pt}}
        \caption{Testing}
        \includegraphics[width=\linewidth]{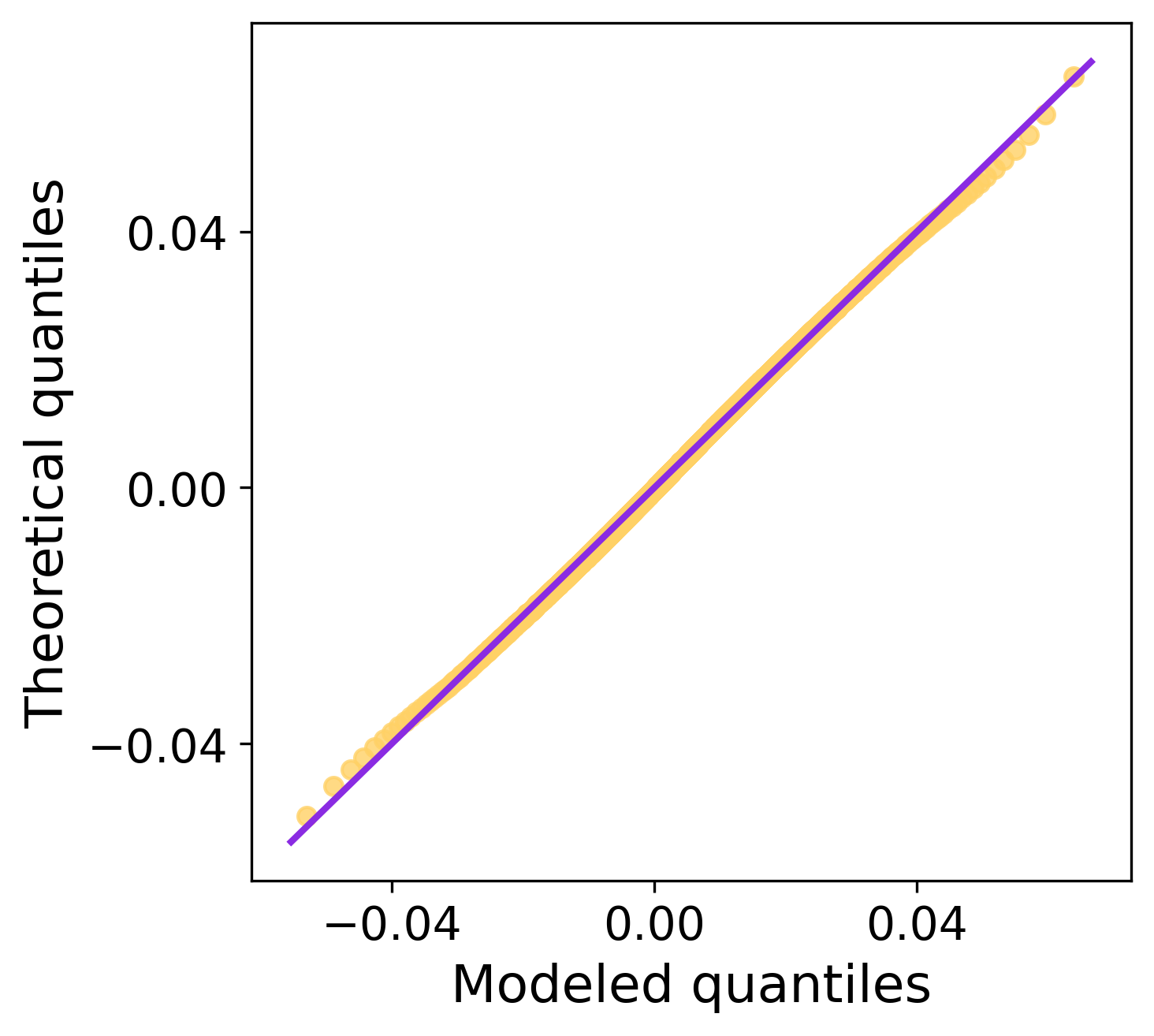}
    \end{subfigure}
    \hspace*{\fill}
    \caption{Quantile--Quantile plots of fitting Gaussian models to dark frames residuals.}
    \label{fig4:dark_qq}
\end{figure}
\subsubsection{Dark-Current Noise}

Dark current arises from thermally generated electrons that accumulate in CCD pixels even without incident light. Since it is thermally activated, its level depends strongly on detector temperature and is substantially reduced by cooling \cite{howell2006handbook,widenhorn2002temperature}. In typical astronomical observations, the detector temperature is regulated and calibration frames are acquired under the same operating conditions as science images \cite{uiowa_ccd_noise}. We therefore treat the dark current as constant for a given instrument setting and neglect explicit temperature dependence in our model.

Following the conventional procedure \cite{mccully2018real}, we represent the dark current as a per-pixel rate in electrons per second, $D_{\rm dark}(x)$ (e$^-$/s), which is retrieved from long-exposure dark frames (300 s) and normalized by the integration time \cite{mccully2018real}. For a science exposure time $t$, the accumulated dark signal can be estimated as:
\begin{equation}
    N_{\rm dark}(x) \sim t\times D_{\rm dark}(x).
\end{equation}
To model $D_{\rm dark}(x)$, we first compute a reference map $D_{\rm ref}(x)$ by averaging the empirically captured dark-current maps $\{D_{\rm dark}^i(x)\}$. We then analyze the sample-to-sample residual $\epsilon_{\rm dark}^i(x)$
and fit a Gaussian distribution to the residuals as:
\begin{equation}
\epsilon_{\rm dark}^i(x) \sim \mathcal{N}\!\left(0,\; \mathrm{Var}_{i,x}[\epsilon_{\rm dark}^i(x)]\right), ~\text{where}~~ \epsilon_{\rm dark}^i(x) = D_{\rm dark}^i(x) - D_{\rm ref}(x)
\end{equation}
This formulation models the spatial pattern of dark-current noise $N_{\rm dark}$ using reference $D_{\rm ref}(x)$, residual $\epsilon_{\rm dark}$, and target exposure duration $t$. The Quantile--Quantile plots for both the fitting set and testing set for $\epsilon_{\rm dark}$ are shown in \Cref{fig4:dark_qq}, suggesting that using a simple Gaussian distribution is sufficient.

\begin{figure}[!tbp]
\centering
\setlength{\tabcolsep}{0pt}
\begin{tabular}{@{} m{0.04\columnwidth} *{3}{m{0.25\columnwidth}} @{}}
  & \multicolumn{3}{@{}l@{}}{%
    \includegraphics[width=0.4\linewidth]{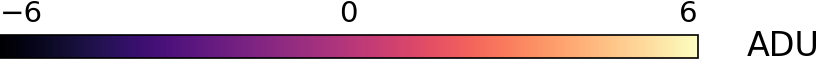}%
  } \\
  \centering\rotatebox{90}{\scriptsize $\topprime{G}$ band} &
  \subcaptionbox{Base Fixed Pattern\label{fig5:bias:a}}[0.244\columnwidth]{%
    \includegraphics[width=\linewidth]{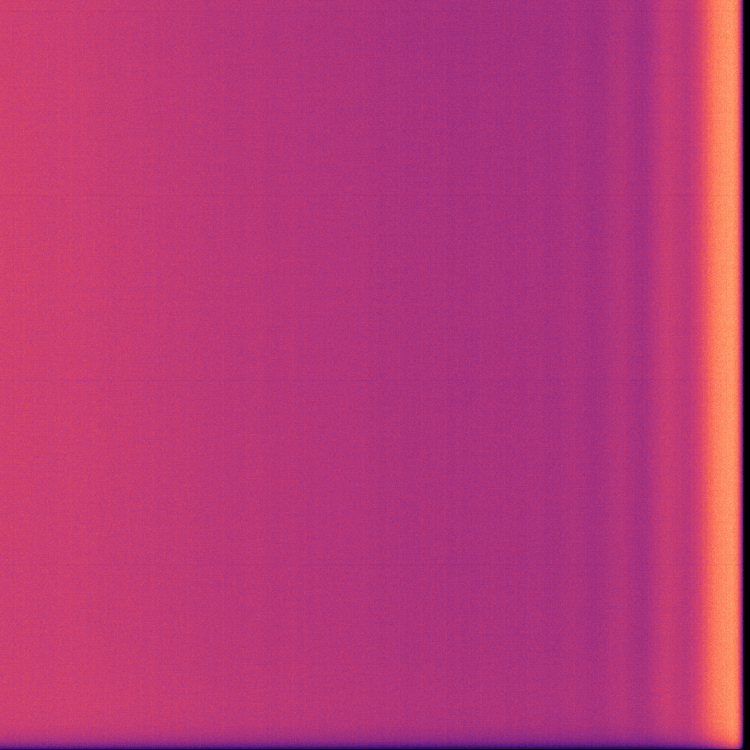}} &
  \subcaptionbox{Synthesized\label{fig5:bias:b}}[0.244\columnwidth]{%
    \includegraphics[width=\linewidth]{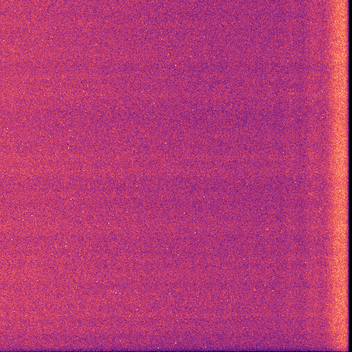}} &
  \subcaptionbox{Real Bias Frame\label{fig5:bias:c}}[0.244\columnwidth]{%
    \includegraphics[width=\linewidth]{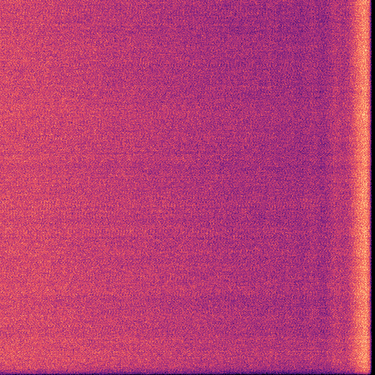}}
\end{tabular}
\caption{Fixed-pattern noise and read noise within 97\% sigma-clipping.}
\label{fig5:bias_synth}
\end{figure}

\subsection{Signal-Independent Noise}
\label{sec:signal_independent_noise}

\subsubsection{Fixed-Pattern Noise and Read Noise}
A CCD bias frame is acquired with zero integration time and records the electronic offset added by the readout chain. Stacking multiple bias frames reduces stochastic read noise and makes deterministic spatial structures more apparent, including \emph{column-wise banding} where certain columns exhibit a consistently elevated bias level \cite{astropy2024guide,pan2025physics}, as shown in the right columns of \Cref{fig5:bias:a}. This banding is a fixed-pattern artifact associated with CCD readout, where charge is transferred through a serial register and measured by one or a few shared output amplifiers, so small baseline drifts or electronic pickup can manifest as coherent column-dependent offsets \cite{astropy2024guide,pan2025physics}.

Given a set of bias frames $\{B_{\rm bias}^i(x)\}$, we first estimate a fixed-pattern template $B_{\rm fp}(x)$ by stacking. This template preserves the stable bias structure, including vertical banding and repeatable offset patterns. For each sample, we compute the residual by subtracting $B_{\rm fp}(x)$, then extract a row-wise bias striping by taking the mean residual within each image row. The remaining residual models pixel-wise readout fluctuations. We fit Gaussian mixture models to both components following \cite{li2025noise} as:
\begin{equation}
N_{\rm row}(\rho(x))\sim\sum_{k=1}^{K_r}\pi_k^{(r)}\,\mathcal{N}_k,\quad
\epsilon_{\rm pix}^i(x)\sim\sum_{k=1}^{K_p}\pi_k^{(p)}\,\mathcal{N}_k,
\end{equation}
where $\rho(x)$ denotes the row index of pixel $x$, $K_r$ and $K_p$ are the numbers of mixture components, and $\{\pi_k^{(\cdot)}\}$ are nonnegative mixture weights that sum to $1$. Each $\mathcal{N}_k$ denotes a Gaussian model.

\begin{figure}[!tp]
    \centering
    \hspace*{\fill}
    \begin{subfigure}[t]{0.32\columnwidth}
        \centering
        \captionsetup{margin={0em,0pt}}
        \caption{Row-pattern}
        \includegraphics[width=\linewidth]{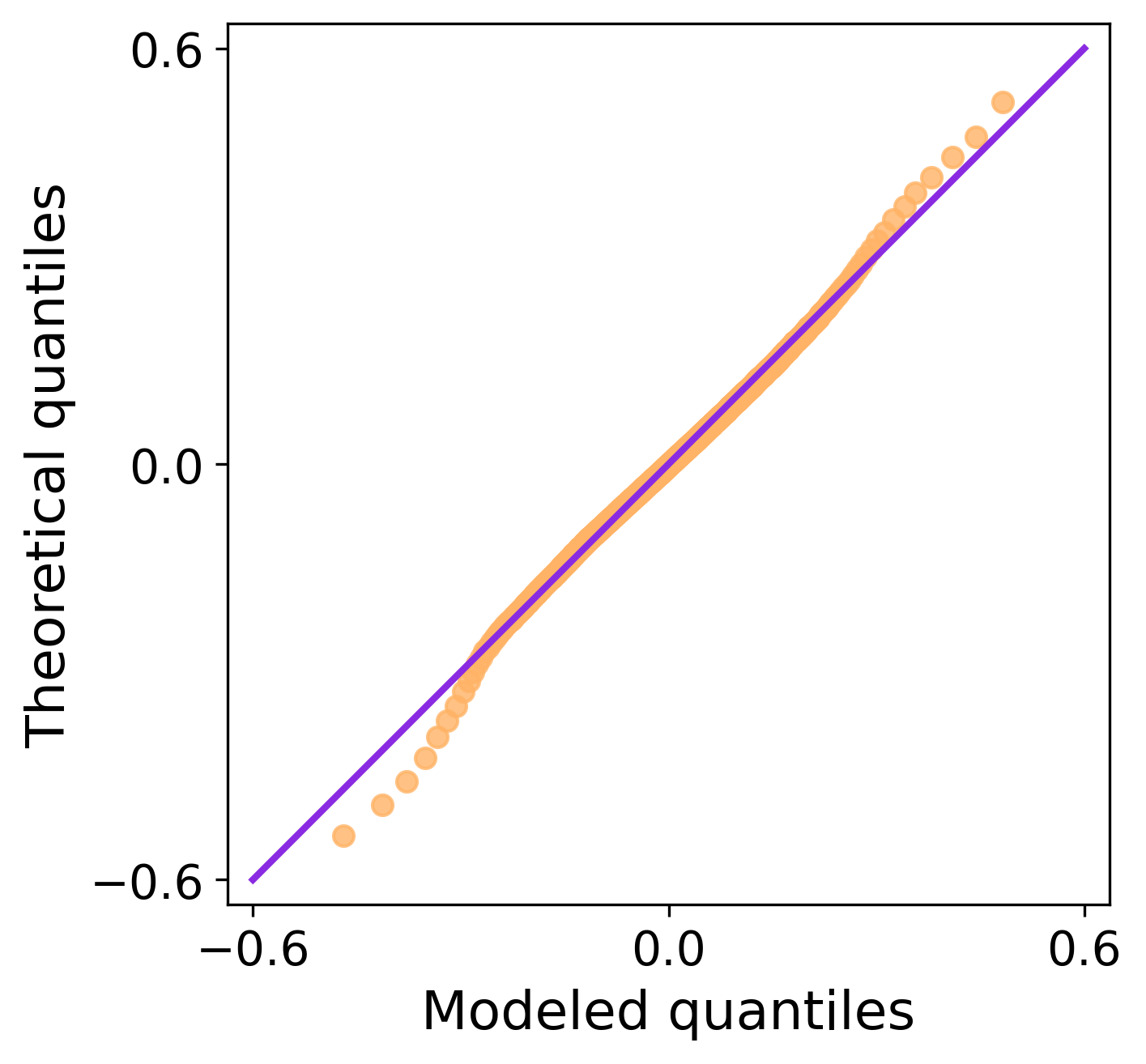}
    \end{subfigure}
    \hspace{0.04\columnwidth}
    \begin{subfigure}[t]{0.32\columnwidth}
        \centering
        \captionsetup{margin={0em,0pt}}
        \caption{Pixel-level}
        \includegraphics[width=\linewidth]{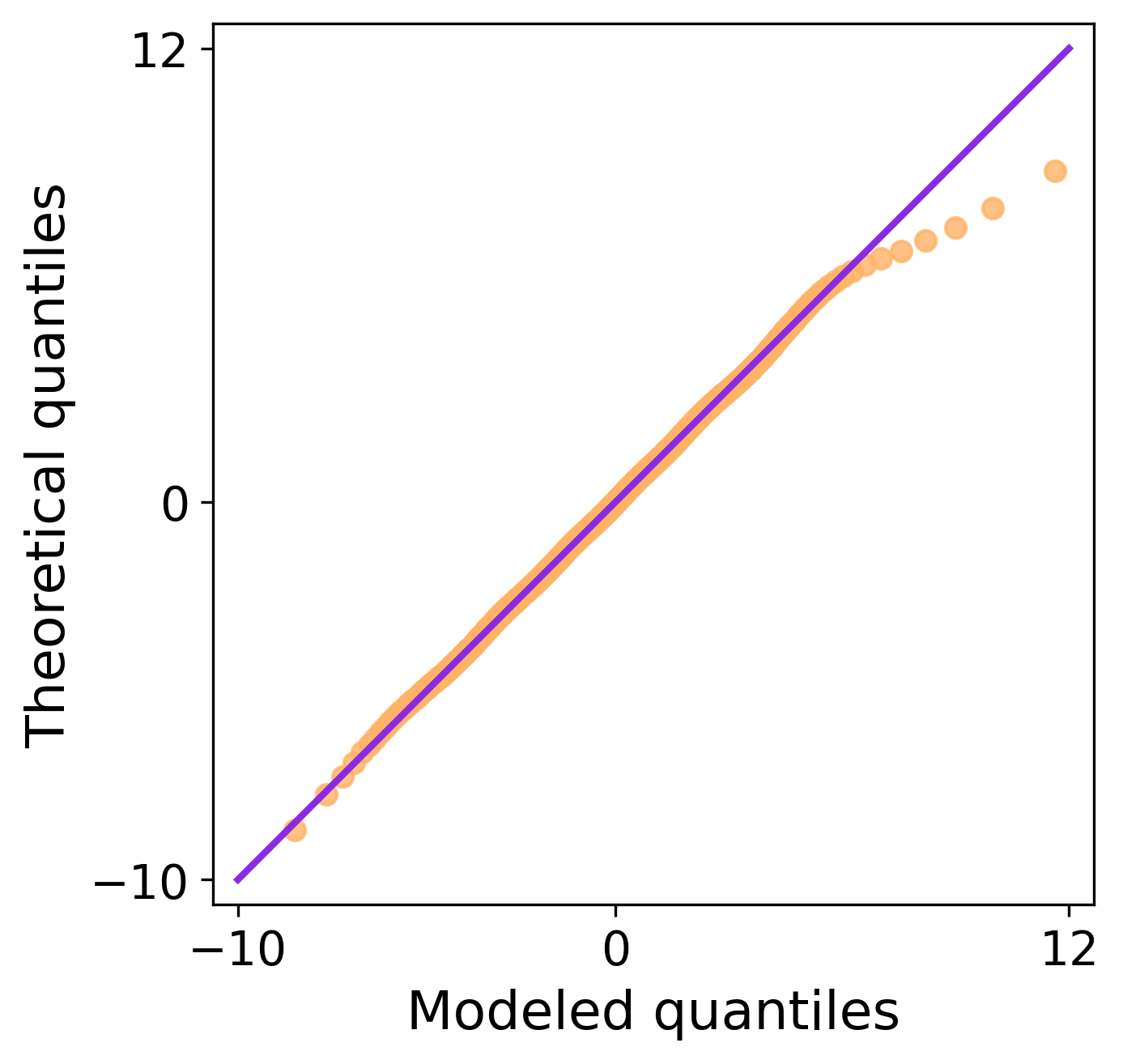}
    \end{subfigure}
    \hspace*{\fill}
    \caption{Quantile–Quantile plots of fitting GMM models to row-pattern and pixel residuals.}
    \label{fig6:bias_qq}
\end{figure}

During synthesis, we sample row offset from the row mixture for each row and broadcast it to all pixels in that row via $N_{\rm row}(\rho(x))$, and sample pixel-wise fluctuations $\epsilon_{\rm pix}(x)$ independently from the pixel mixture. The resulting read-noise realization is constructed as:
\begin{equation}
N_{\rm read}(x)=B_{\rm fp}(x)+N_{\rm row}(\rho(x))+\epsilon_{\rm pix}(x),
\end{equation}
which preserves the observed fixed-pattern bias structure through $B_{\rm fp}(x)$ while reproducing realistic row-level and pixel-level read-noise variability. \Cref{fig6:bias_qq} shows the testing results of the Quantile--Quantile plot, with a close fit to both theoretical distributions. 

\subsubsection{Digitization Error}

After charge-to-voltage conversion, the CCD output is digitized by an analog-to-digital converter, mapping a continuous signal to a discrete digital number (ADU). This introduces a small quantization error \(N_{\rm dig}(x)\sim U(-\Delta/2,\Delta/2)\), where \(\Delta=G\) in electron units. Because read noise is empirically estimated from bias frames, this digitization error is already absorbed into the measured read-noise term \cite{howell2006handbook}, so we do not model \(N_{\rm dig}(x)\) separately.


\subsubsection{Hot-Pixel and Cosmic Ray}
Hot pixels arise from detector defects with abnormally high dark current. Their intensities increase approximately linearly with exposure time due to dark-signal accumulation, and their locations remain fixed across frames \cite{stsi2024wfc3}. Cosmic rays are transient high-energy events that deposit charge in the sensor and appear as impulsive bright outliers. In ground-based CCD imaging, cosmic-ray hits typically contaminate a small fraction of pixels per long exposure, affecting approximately $0.01\%$--$0.1\%$ of image pixels \cite{groom2002cosmic,farage2005evaluation,lupton2005sdss}.

Rather than counting persistent hot pixels and transient cosmic rays separately, we model both effects using a single salt-and-pepper impulse noise model:
\begin{equation}
N_{\rm hot}(x)=Z(x)\,A(x),\qquad
Z(x)\sim{\rm Bernoulli}\!\left(p_{\rm hot}\right),
\end{equation}
where $A(x)$ is drawn from a uniform distribution $[5000, 10000]$, and $p_{\rm hot}=0.01\%$ denotes the impulse probability that controls the fraction of corrupted pixels.

\section{MuSCAT Observer Dataset}
\label{sec:dataset}

\subsubsection{Instruments}
MuSCAT-3 (Haleakalā, Hawaii; LCO-FTN, 2020) and MuSCAT-4 (Siding Spring, Australia; LCO-FTS, 2023) are twin four-channel optical imagers installed on 2\,m Las Cumbres Observatory telescopes. Each instrument uses a cascade of dichroic beam splitters to feed four independent CCD cameras in the $\topprime{G}$, $\topprime{R}$, $\topprime{I}$, and $Z_s$ bands. In this work, we focus on $\topprime{G}$, $\topprime{R}$, and $\topprime{I}$, which follow the SDSS filter system and are the most consistently calibrated and frequently observed channels. Since MuSCAT-3 and MuSCAT-4 share nearly identical opto-mechanical designs and the same filter set, they enable cross-instrument zero-shot evaluation; we treat each band as an independent dataset to assess denoising performance under distinct, device-dependent noise characteristics.

\subsubsection{Data Statistics}
\label{sec:data_stats}
We construct MuSCAT-3 for training and in-domain testing, and reserve MuSCAT-4 for zero-shot evaluation. For each observation, we provide (i) the raw science frame, (ii) the BANZAI ``processed'' product, and (iii) the calibration frames, including bias, dark, and skyflats. BANZAI is the official LCO real-time reduction pipeline and performs standard instrumental signature removal \cite{mccully2018real}. \Cref{tab:data_stats} summarizes the dataset statistics per band.

While BANZAI removes deterministic detector and optical artifacts, the calibrated science images still largely contain photon-shot noise \cite{robberto2014refpix_nircam,mccully2018real}. To obtain high-SNR base images for noise synthesis, we additionally collect 10--20 consecutive BANZAI-processed exposures for each science target. We stack these exposures to form a high-SNR reference image \cite{fischer1993optimal,craig2023image}, denoted as \texttt{Mean}. \Cref{tab:noise_stats} reports background noise levels measured on source-masked pixels using both normalized median absolute deviation (NMAD) and standard deviation (STD).

\begin{table*}[!tp]
\centering
\setlength{\tabcolsep}{3pt}
\scriptsize
\begin{minipage}[t]{0.49\textwidth}
\centering
\captionof{table}{Dataset statistics of MuSCAT-3 and MuSCAT-4 datasets.}
\label{tab:data_stats}
\begin{tabularx}{\linewidth}{l|cccccc}
\toprule
& \multicolumn{3}{c}{\textbf{MuSCAT-3}} & \multicolumn{3}{c}{\textbf{MuSCAT-4}} \\
\cmidrule(lr){2-4}\cmidrule(lr){5-7}
\textbf{Data} & $\topprime{G}$ & $\topprime{I}$ & $\topprime{R}$ & $\topprime{G}$ & $\topprime{I}$ & $\topprime{R}$ \\
\midrule
Training     & 215 & 435 & 395 & - & - & - \\
Testing      & 54 & 73 & 61 & 46 & 96 & 72 \\
Calibration  & 807 & 1524 & 1368 & 138 & 288 & 216 \\
Processed    & 4843 & 9578 & 9677 & 661 & 1687 & 1566 \\
\bottomrule
\end{tabularx}
\end{minipage}
\hfill
\begin{minipage}[t]{0.49\textwidth}
\setlength{\tabcolsep}{4pt}
\centering
\captionof{table}{Comparison of background noise levels under different processing methods.}
\label{tab:noise_stats}
\begin{tabularx}{\linewidth}{l|cccc}
\toprule
& \multicolumn{2}{c}{\textbf{MuSCAT-3}} & \multicolumn{2}{c}{\textbf{MuSCAT-4}} \\
\cmidrule(lr){2-3}\cmidrule(lr){4-5}
\textbf{Methods} & NMAD$\downarrow$ & STD$\downarrow$ & NMAD$\downarrow$ & STD$\downarrow$ \\
\midrule
Raw     & 10.50 & 11.73 & 11.61 & 11.73 \\
BANZAI  & 9.33  & 9.54  & 9.40  & 9.63  \\
\textbf{Mean}  & 3.12  & 4.72  & 3.70  & 4.72  \\
Drizzle & 6.15  & 6.48  & 6.22  & 6.55  \\
\bottomrule
\end{tabularx}
\end{minipage}
\end{table*}

\section{Experiment}
\label{sec:experiment}

\subsubsection{Implementation Details}
Following previous studies \cite{chen2018learning,wei2021physics,li2025noise}, we construct a neural denoising pipeline using a standard U-Net architecture \cite{ronneberger2015u}. We use high-SNR stacking-mean images as clean baselines and synthesize noisy observations to form paired training data. We sample 100 calibration frames from the training set to build our physics-based noise model. We train the U-Net with 140 epochs using Raw-L1 loss and Adam optimizer with batch size 8 on a single H200 GPU. The learning rate is $1\times10^{-4}$. Detailed information is given in Appendix D.

\subsubsection{Baseline Methods}
We evaluate both traditional and learning-based baselines.
Traditional methods include BM3D \cite{dabov2007image} and A-BM3D \cite{tibbs2018denoising} for single-image denoising, and Drizzle \cite{fruchter2002drizzle} for multi-image reconstruction from aligned exposures.
Learning-based baselines include supervised methods trained with direct paired data or PMN \cite{feng2023learnability}, as well as self-supervised approaches including Noise2Noise (N2N) \cite{lehtinen2018noise}, ELD \cite{cao2023physics}, SFRN \cite{zhang2021rethinking}, PNNP \cite{feng2023physics}, and NMOH \cite{li2025noise}. Implementation details of baseline approaches are outlined in the Appendix E.

\subsubsection{Metric}
For synthetic data generated with our noise formation model, we report PSNR to evaluate photometric fidelity. For real observations, we report FLUX-SNR (dB) to assess scientific accuracy. We also compute NMAD and STD on source-masked background pixels, which are expected to be near constant, to quantify residual noise levels. We clip both prediction and GT at 5000, corresponding to 99.99\% of pixels, to suppress extreme outliers that could bias photometric evaluation. Best results are shaded as \colorbox{fst}{first}, \colorbox{sec}{second}, and \colorbox{thd}{third}.

\subsection{Evaluation on MuSCAT-3 Dataset}

\begin{figure}[t]
    \begin{center}
        \includegraphics[width=\textwidth]{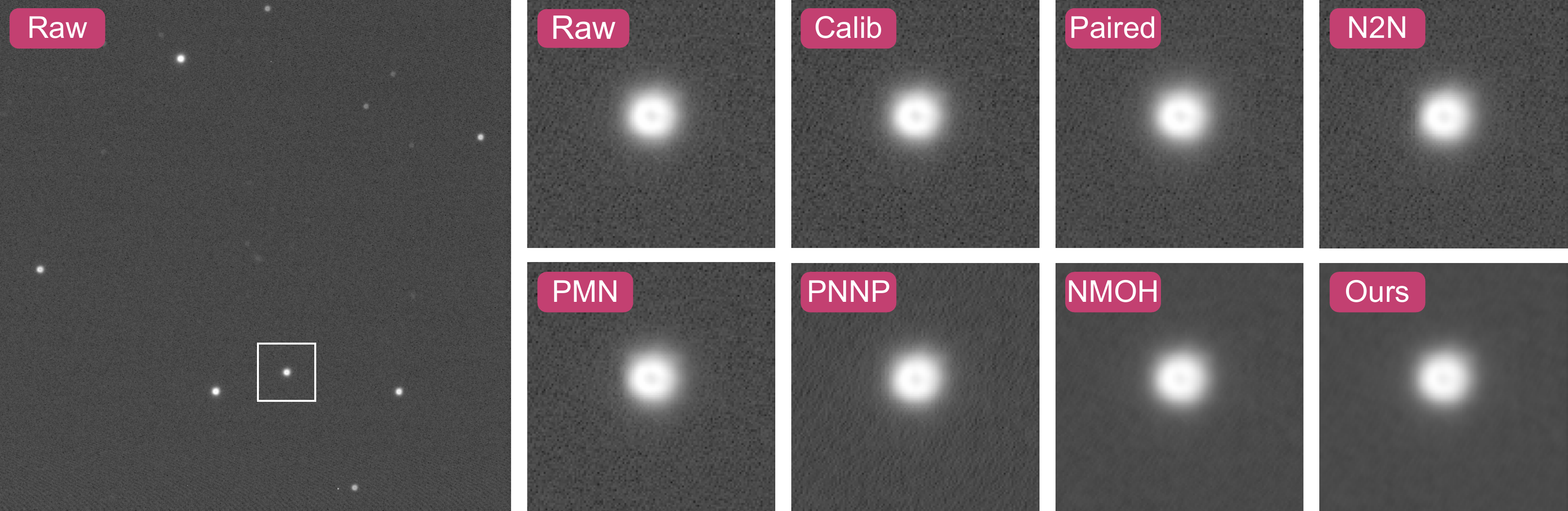}
    \end{center}
    \caption{Visualization of baseline method comparisons on the MuSCAT-3 dataset, $\topprime{G}$ band, targeting TOI-4643 (2022-10-26 observation). Zoom in for better visibility.}
    \label{fig:exp_muscat3}
\end{figure}

\begin{figure}[t]
    \begin{center}
        \includegraphics[width=\textwidth]{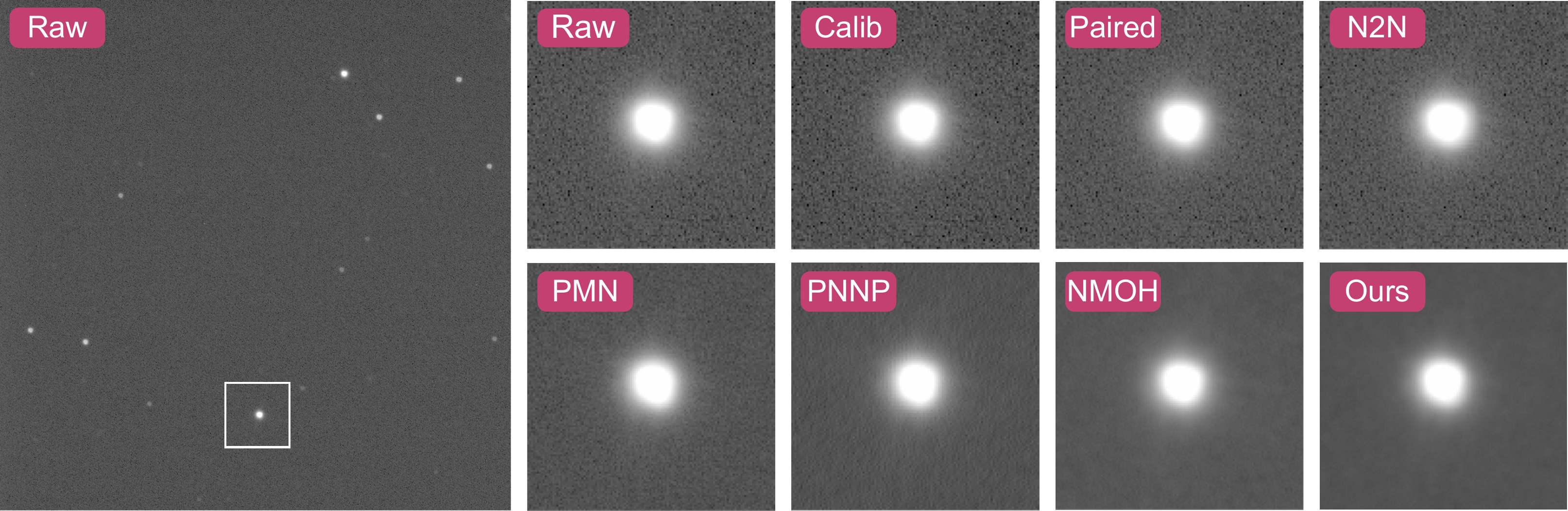}
    \end{center}
    \caption{Visualization of zero-shot evaluation on the MuSCAT-4 dataset, $\topprime{R}$ band, targeting WASP-53 (2023-10-08 observation). Zoom in for better visibility.}
    \label{fig:exp_muscat4}
\end{figure}

\Cref{tab:muscat3} reports quantitative results on the MuSCAT-3 dataset for both synthetic and real observations across three bands. For the synthetic setting, we use our noise formation model to generate realistic noisy raw observations from stacked clean proxies, while incorporating the corresponding per-observation calibration frames acquired on the same night to match real acquisition conditions. Under this synthetic evaluation, our method consistently outperforms all baselines and achieves the best photometric accuracy.

To evaluate scientific accuracy on real observations, where GT is typically inaccessible, we follow common practice in astronomy by injecting mock PSFs \cite{guo2026deeper} with predefined photon flux into real observations and computing FLUX-SNR against the mocked GT. In addition, we report background noise statistics, including NMAD and STD, measured on source-masked regions, for a complementary evaluation of background smoothness and residual noise suppression. Compared with learning-based approaches \cite{feng2023learnability,lehtinen2018noise,cao2023physics,zhang2021rethinking,feng2023physics,li2025noise}, our method consistently achieves superior flux accuracy and background smoothness, whereas baseline methods designed for commercial cameras often overlook PRNU and CCD-specific noise characteristics. Compared with traditional denoisers \cite{dabov2007image,tibbs2018denoising,fruchter2002drizzle}, although they may achieve stronger noise suppression, reflected by lower NMAD and STD, they tend to oversmooth stellar PSFs, attenuating peak intensities and fine PSF structures, which leads to lower FLUX-SNR.

\begin{table*}[t]
\centering
\caption{Quantitative evaluation on MuSCAT-3 dataset.}
\setlength{\tabcolsep}{0pt}
\renewcommand{\arraystretch}{1.15}
\fontsize{6pt}{7pt}\selectfont
\begin{tabularx}{\textwidth}{
    @{\hspace{1pt}}l@{\hspace{3pt}} |
    @{\hspace{2pt}}l@{\hspace{2pt}} |
    *{4}{>{\centering\arraybackslash}X} |
    *{4}{>{\centering\arraybackslash}X} |
    *{4}{>{\centering\arraybackslash}X} |
    *{4}{>{\centering\arraybackslash}X}
}
\toprule
&  & \multicolumn{4}{c|}{PSNR(Syn)$\uparrow$}
   & \multicolumn{4}{c|}{FLUX-SNR(Real)$\uparrow$}
   & \multicolumn{4}{c|}{NMAD(Real)$\downarrow$}
   & \multicolumn{4}{c}{STD(Real)$\downarrow$} \\
\cmidrule(lr){3-6}\cmidrule(lr){7-10}\cmidrule(lr){11-14}\cmidrule(lr){15-18}
& \textbf{Method}
& $\topprime{G}$ & $\topprime{I}$ & $\topprime{R}$ & \textbf{Avg}
& $\topprime{G}$ & $\topprime{I}$ & $\topprime{R}$ & \textbf{Avg}
& $\topprime{G}$ & $\topprime{I}$ & $\topprime{R}$ & \textbf{Avg}
& $\topprime{G}$ & $\topprime{I}$ & $\topprime{R}$ & \textbf{Avg} \\
\midrule
\multirow{3}{*}{\rotatebox[origin=c]{90}{Trad}}
& BM3D\cite{dabov2007image}
& 55.03 & 54.88 & 54.79 & 54.90
& 37.56 & \cellcolor{thd}38.28 & 37.10 & \cellcolor{thd}37.64
& 2.59 & 3.11 & 2.58 & 2.76
& \cellcolor{thd}3.21 & 3.31 & \cellcolor{thd}2.99 & \cellcolor{thd}3.17 \\
& ABM3D\cite{tibbs2018denoising}
& 55.38 & 54.57 & 54.60 & 54.85
& \cellcolor{sec}39.48 & \cellcolor{sec}38.44 & \cellcolor{fst}40.13 & \cellcolor{sec}39.35
& \cellcolor{fst}1.20 & \cellcolor{fst}1.15 & \cellcolor{fst}1.15 & \cellcolor{fst}1.17
& \cellcolor{fst}2.39 & \cellcolor{fst}2.13 & \cellcolor{fst}2.11 & \cellcolor{fst}2.21 \\
& Drizzle\cite{fruchter2002drizzle}
& - & - & - & -
& - & - & - & -
& 6.05 & 6.08 & 6.31 & 6.15
& 6.36 & 6.50 & 6.58 & 6.48 \\
\midrule
\rowcolor{gray!15}
\cellcolor{white}%
\multirow{8}{*}{\rotatebox[origin=c]{90}{Learning}}
& Paired
& 54.74 & 56.07 & 56.22 & 55.68
& 26.13 & 28.86 & 27.14 & 27.38
& 7.56 & 9.28 & 9.02 & 8.62
& 7.84 & 9.50 & 9.21 & 8.85 \\
& PMN\cite{feng2023learnability}
& \cellcolor{thd}57.06 & \cellcolor{thd}56.68 & 56.76 & \cellcolor{thd}56.83
& 30.55 & 30.48 & 29.37 & 30.13
& 8.13 & 9.34 & 8.95 & 8.81
& 8.35 & 9.48 & 9.15 & 8.99 \\
& N2N\cite{lehtinen2018noise}
& 56.29 & 56.66 & \cellcolor{thd}56.91 & 56.62
& 36.16 & 36.11 & 35.96 & 36.08
& 7.76 & 9.29 & 8.92 & 8.66
& 8.02 & 9.46 & 9.12 & 8.87 \\
& ELD\cite{cao2023physics}
& 56.05 & 55.89 & 56.38 & 56.11
& 32.16 & 30.02 & 28.12 & 30.10
& 2.60 & \cellcolor{thd}2.32 & 2.58 & 2.50
& 3.48 & 3.58 & 4.49 & 3.85 \\
& SFRN\cite{zhang2021rethinking}
& 51.23 & 56.13 & 55.70 & 54.35
& 28.87 & 29.47 & 28.39 & 28.91
& 4.20 & 4.98 & 4.82 & 4.67
& 5.27 & 5.75 & 5.55 & 5.52 \\
& PNNP\cite{feng2023physics}
& \cellcolor{sec}57.10 & \cellcolor{sec}56.88 & 56.80 & \cellcolor{sec}56.92
& \cellcolor{thd}38.16 & 37.52 & \cellcolor{thd}37.11 & 37.60
& 6.39 & 7.95 & 7.64 & 7.33
& 6.75 & 8.04 & 7.73 & 7.51 \\
& NMOH\cite{li2025noise}
& 56.03 & 56.28 & \cellcolor{thd}56.86 & 56.39
& 37.51 & 36.26 & 36.83 & 36.87
& \cellcolor{thd}2.58 & 2.35 & \cellcolor{thd}2.45 & \cellcolor{thd}2.46
& 3.46 & \cellcolor{thd}3.21 & 3.45 & 3.37 \\
& \textbf{Ours}
& \cellcolor{fst}57.29 & \cellcolor{fst}56.95 & \cellcolor{fst}57.09 & \cellcolor{fst}57.06
& \cellcolor{fst}41.21 & \cellcolor{fst}38.50 & \cellcolor{sec}40.03 & \cellcolor{fst}39.91
& \cellcolor{sec}2.21 & \cellcolor{sec}2.25 & \cellcolor{sec}2.31 & \cellcolor{sec}2.43
& \cellcolor{sec}2.92 & \cellcolor{sec}2.86 & \cellcolor{thd}3.42 & \cellcolor{sec}3.06 \\
\bottomrule
\end{tabularx}
\label{tab:muscat3}
\end{table*}

\begin{table*}[t]
\centering
\caption{Quantitative zero-shot evaluation on MuSCAT-4 dataset.}
\setlength{\tabcolsep}{0pt}
\renewcommand{\arraystretch}{1.15}
\fontsize{6pt}{7pt}\selectfont
\begin{tabularx}{\textwidth}{
    @{\hspace{1pt}}l@{\hspace{3pt}} |
    @{\hspace{2pt}}l@{\hspace{2pt}} |
    *{4}{>{\centering\arraybackslash}X} |
    *{4}{>{\centering\arraybackslash}X} |
    *{4}{>{\centering\arraybackslash}X} |
    *{4}{>{\centering\arraybackslash}X}
}
\toprule
&  & \multicolumn{4}{c|}{PSNR(Syn)$\uparrow$}
   & \multicolumn{4}{c|}{FLUX-SNR(Real)$\uparrow$}
   & \multicolumn{4}{c|}{NMAD(Real)$\downarrow$}
   & \multicolumn{4}{c}{STD(Real)$\downarrow$} \\
\cmidrule(lr){3-6}\cmidrule(lr){7-10}\cmidrule(lr){11-14}\cmidrule(lr){15-18}
& \textbf{Method}
& $\topprime{G}$ & $\topprime{I}$ & $\topprime{R}$ & \textbf{Avg}
& $\topprime{G}$ & $\topprime{I}$ & $\topprime{R}$ & \textbf{Avg}
& $\topprime{G}$ & $\topprime{I}$ & $\topprime{R}$ & \textbf{Avg}
& $\topprime{G}$ & $\topprime{I}$ & $\topprime{R}$ & \textbf{Avg} \\
\midrule
\multirow{3}{*}{\rotatebox[origin=c]{90}{Trad}}
& BM3D\cite{dabov2007image}
& 55.07 & 55.55 & 56.61 & 55.74
& \cellcolor{thd}39.00 & \cellcolor{thd}37.56 & \cellcolor{thd}39.13 & \cellcolor{thd}38.56
& 2.99 & \cellcolor{fst}2.34 & \cellcolor{sec}2.26 & \cellcolor{thd}2.53
& 4.04 & \cellcolor{thd}3.96 & \cellcolor{sec}3.62 & \cellcolor{thd}3.87 \\
& ABM3D\cite{tibbs2018denoising}
& 57.75 & 57.25 & 56.92 & 57.31
& \cellcolor{sec}40.51 & \cellcolor{sec}37.70 & \cellcolor{fst}40.04 & \cellcolor{sec}39.42
& \cellcolor{sec}{2.42} & \cellcolor{sec}{2.45} & \cellcolor{fst}{2.18} & \cellcolor{fst}{2.35}
& \cellcolor{thd}{3.69} & \cellcolor{fst}{3.51} & \cellcolor{fst}{3.37} & \cellcolor{fst}{3.52} \\
& Drizzle\cite{fruchter2002drizzle}
& - & - & - & -
& - & - & - & -
& 6.17 & 6.51 & 5.98 & 6.22
& 6.24 & 6.90 & 6.51 & 6.55 \\
\midrule
\rowcolor{gray!15}
\cellcolor{white}%
\multirow{8}{*}{\rotatebox[origin=c]{90}{Learning}}
& Paired
& 57.11 & 56.52 & 56.63 & 56.75
& 25.33 & 28.36 & 29.12 & 27.60
& 8.11 & 10.06 & 9.12 & 9.10
& 8.49 & 10.57 & 9.36 & 9.47 \\
& PMN\cite{feng2023learnability}
& \cellcolor{thd}{59.13} & 56.39 & 56.62 & 57.38
& 29.88 & 30.97 & 29.64 & 30.16
& 8.77 & 10.12 & 9.05 & 9.31
& 8.99 & 10.63 & 9.38 & 9.67 \\
& N2N\cite{lehtinen2018noise}
& 58.21 & 56.75 & 57.25 & 57.40
& 37.03 & 36.51 & 36.06 & 36.53
& 8.42 & 10.10 & 9.10 & 9.21
& 8.70 & 10.59 & 9.31 & 9.53 \\
& ELD\cite{cao2023physics}
& \cellcolor{thd}{59.13} & 57.29 & 56.81 & 57.74
& 32.56 & 30.65 & 31.04 & 31.42
& 2.87 & 2.59 & \cellcolor{thd}{2.53} & 2.66
& \cellcolor{sec}{3.49} & 4.33 & \cellcolor{thd}{3.81} & 3.88 \\
& SFRN\cite{zhang2021rethinking}
& 53.80 & 53.33 & 52.08 & 53.07
& 28.17 & 29.80 & 29.55 & 29.17
& 4.50 & 5.01 & 4.47 & 4.66
& 5.12 & 6.32 & 5.58 & 5.67 \\
& PNNP\cite{feng2023physics}
& \cellcolor{sec}{59.30} & \cellcolor{thd}{57.36} & \cellcolor{sec}{57.72} & \cellcolor{thd}{58.13}
& 37.76 & 36.27 & 37.07 & 37.03
& 7.15 & 7.60 & 6.85 & 7.20
& 7.56 & 8.27 & 7.40 & 7.74 \\
& NMOH\cite{li2025noise}
& 59.04 & \cellcolor{sec}{57.93} & \cellcolor{thd}{57.46} & \cellcolor{sec}{58.14}
& 36.95 & 36.28 & 36.34 & 36.52
& \cellcolor{thd}{2.79} & 2.68 & 2.54 & {2.67}
& 3.80 & 4.49 & 4.07 & 4.12 \\
& \textbf{Ours}
& \cellcolor{fst}{60.22} & \cellcolor{fst}{58.34} & \cellcolor{fst}{57.98} & \cellcolor{fst}{58.85}
& \cellcolor{fst}42.07 & \cellcolor{fst}37.98 & \cellcolor{sec}39.56 & \cellcolor{fst}39.87
& \cellcolor{fst}{2.35} & \cellcolor{thd}{2.51} & \cellcolor{thd}{2.53} & \cellcolor{sec}{2.49}
& \cellcolor{fst}{3.48} & \cellcolor{sec}{3.83} & \cellcolor{sec}{3.62} & \cellcolor{sec}{3.64} \\
\bottomrule
\end{tabularx}
\label{tab:muscat4}
\end{table*}

\subsection{Zero-shot Evaluation on MuSCAT-4 Dataset}
We further conduct a cross-instrument evaluation by applying the model trained on MuSCAT-3 observations to denoise MuSCAT-4 in a zero-shot manner, as reported in \Cref{tab:muscat4}. The results follow trends consistent with \Cref{tab:muscat3}, validating the generalization ability of our physics-based training pipeline across instruments and observing conditions.

Moreover, \Cref{fig:exp_muscat3} and \Cref{fig:exp_muscat4} provide qualitative comparisons on real $\topprime{G}$-band and $\topprime{I}$-band observations from the MuSCAT-3 and 4 datasets, showing that our method delivers optimal and photometrically faithful denoising performance among baseline approaches while preserving stellar PSF structures.

\section{Ablation Study}
\subsubsection{Effect of Noise Formation Components}

\begin{wraptable}{r}{0.45\columnwidth}
\vspace{-2.85\baselineskip}
\centering
\fontsize{6pt}{7pt}\selectfont
\caption{Ablation study on each noise components in \Cref{eq:noise_formation} evaluated on \textbf{\textit{real}} observations with mocked PSF. Left cell: MuSCAT-3 $\topprime{G}$. Right cell: zero-shot on the MuSCAT-4 $\topprime{G}$.}
\label{tab:ablation_noise}
\setlength{\tabcolsep}{1pt}
\renewcommand{\arraystretch}{1}
\begin{tabularx}{0.45\columnwidth}{
  c|l|
  *{5}{>{\centering\arraybackslash}X|}%
  >{\centering\arraybackslash}X
}
\toprule
\multicolumn{2}{l|}{Methods} &
\multicolumn{2}{c|}{FLUX-SNR$\uparrow$} &
\multicolumn{2}{c|}{NMAD$\downarrow$} &
\multicolumn{2}{c}{STD$\downarrow$} \\
\midrule
\multirow{5}{*}{\rotatebox[origin=c]{90}{w/o}}
& PRNU
& 35.51 & 35.17
& 3.02  & 3.26
& 3.76  & 4.08 \\
& $N_{\rm dark}$
& \cellcolor{thd}38.21 & \cellcolor{thd}37.62
& 2.78  & 2.99
& \cellcolor{thd}3.66 & \cellcolor{thd}3.68 \\
& $N_{\rm phot}$
& 32.16 & 31.51
& 5.72  & 5.90
& 5.97  & 6.03 \\
& $N_{\rm read}$
& 31.69 & 32.03
& \cellcolor{fst}{2.38} & \cellcolor{thd}2.98
& 3.81  & 3.83 \\
& $N_{\rm hot}$
& \cellcolor{sec}40.88 & \cellcolor{sec}41.64
& \cellcolor{sec}2.46 & \cellcolor{sec}{2.82}
& \cellcolor{fst}{3.32} & \cellcolor{fst}{3.36} \\
\midrule
\multicolumn{2}{l|}{Ours}
& \cellcolor{fst}41.21 & \cellcolor{fst}42.07
& \cellcolor{thd}2.58 & \cellcolor{fst}2.79
& \cellcolor{sec}3.46 & \cellcolor{sec}3.48 \\
\bottomrule
\end{tabularx}
\end{wraptable}

\begin{figure}[!tp]
    \begin{center}
        \includegraphics[width=\textwidth]{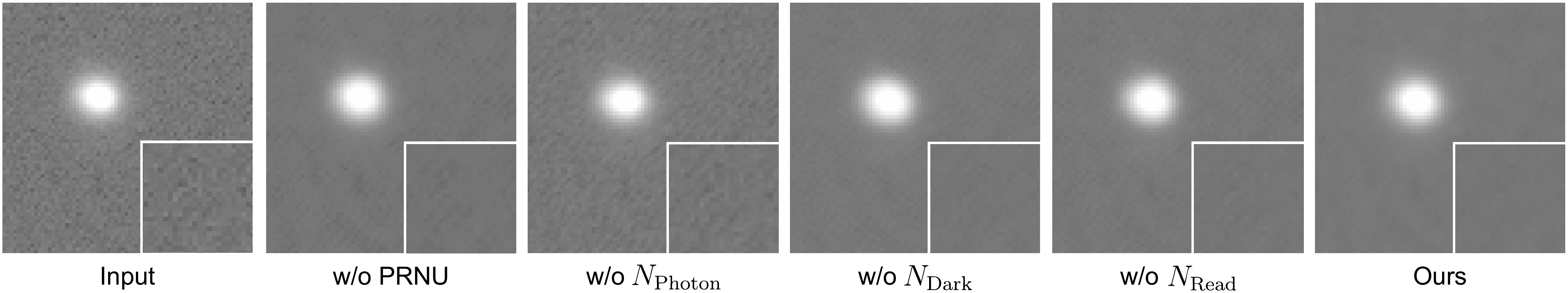}
    \end{center}
    \caption{Visualization of ablation studies on MuSCAT-3 dataset $\topprime{G}$-band, targeting the quasar SDSS J0106 (2021-11-25 observation).}
    \label{fig:ablation_vis}
\end{figure}

We conduct ablation studies by synthesizing training pairs with each component removed at a time and training the networks under the same settings. Quantitative results are reported in \Cref{tab:ablation_noise}, with representative visual comparisons shown in \Cref{fig:ablation_vis}. Specifically, removing $\mathrm{PRNU}$, $N_{\rm phot}$, and $N_{\rm read}$ significantly reduces FLUX-SNR, leading to pronounced scientific artifacts in PSF flux recovery. These failure cases are consistent with \Cref{fig:ablation_vis}, where stronger residual noise remains in the restored results. In particular, $N_{\rm read}$ models fixed-pattern noise, which can substantially affect flux estimation when the pattern overlaps with mocked PSFs. Moreover, removing $N_{\rm hot}$ can produce lower noise levels on source-masked backgrounds, reflected by lower NMAD and STD; however, modeling such artifacts is still essential for optimal flux estimation, as it improves robustness to real impulsive corruptions. Across the remaining ablations, removing any component consistently degrades performance, underscoring the necessity of the full noise formation model.

\begin{figure*}[!tp]
\centering
\captionsetup[sub]{font=scriptsize, labelfont=bf, skip=1pt}
\setlength{\tabcolsep}{2pt}
\begin{tabular}{@{}cccc@{}}
\subcaptionbox{3$\topprime{G}$ (FLUX-SNR$\uparrow$)\label{fig:ablation:num:a}}{\includegraphics[width=0.24\textwidth]{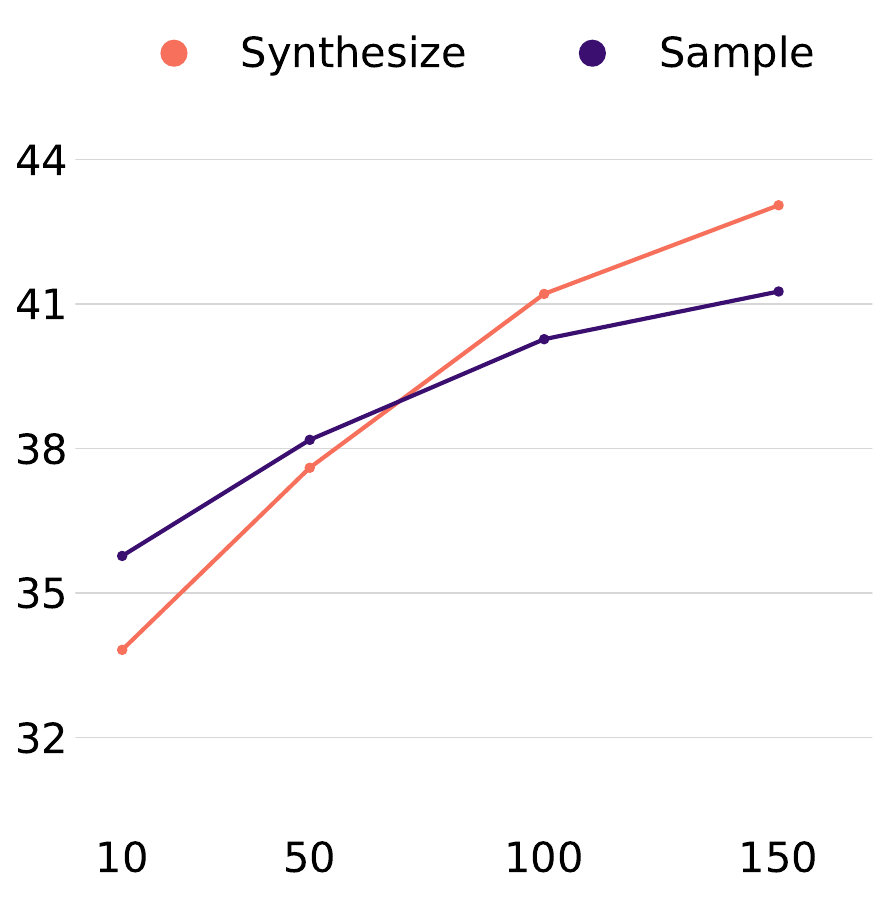}} &
\subcaptionbox{4$\topprime{G}$ (FLUX-SNR$\uparrow$)\label{fig:ablation:num:b}}{\includegraphics[width=0.24\textwidth]{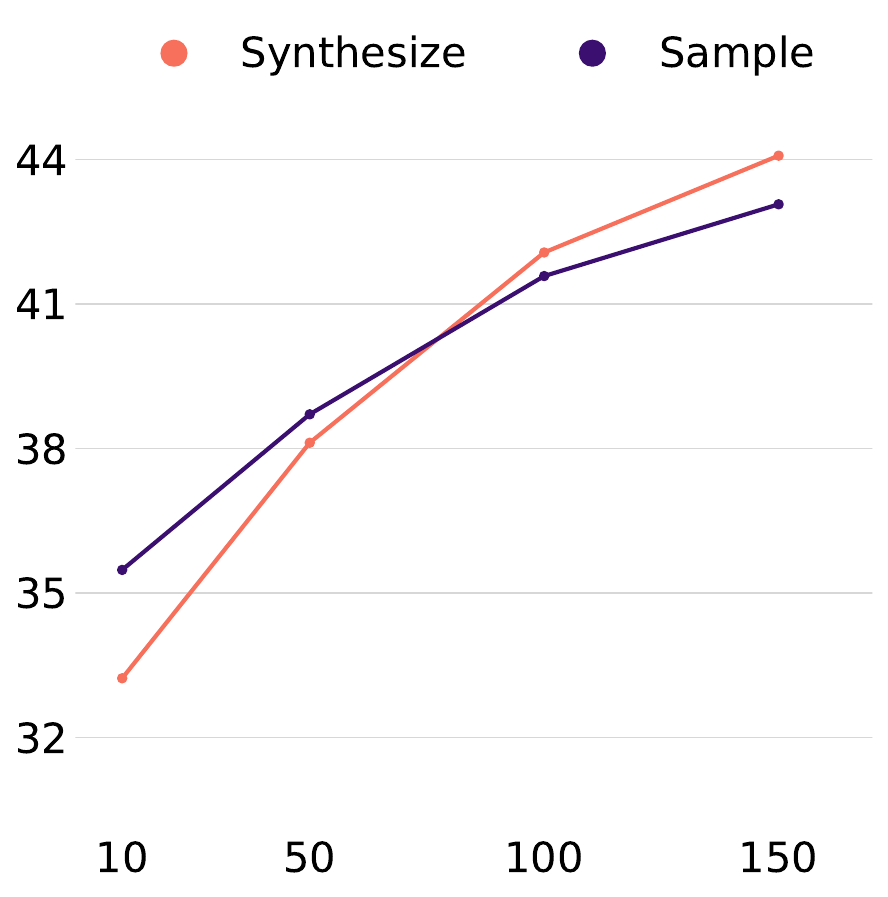}} &
\subcaptionbox{3$\topprime{G}$ (NMAD$\downarrow$)\label{fig:ablation:num:c}}{\includegraphics[width=0.24\textwidth]{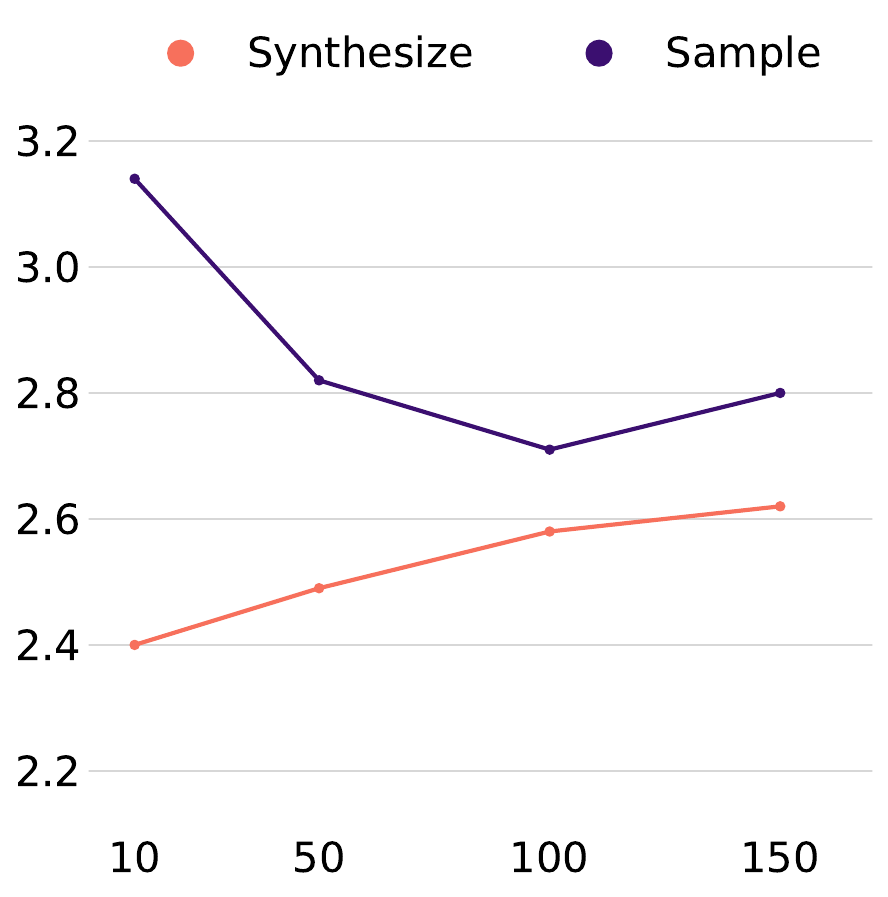}} &
\subcaptionbox{4$\topprime{G}$ (NMAD$\downarrow$)\label{fig:ablation:num:d}}{\includegraphics[width=0.24\textwidth]{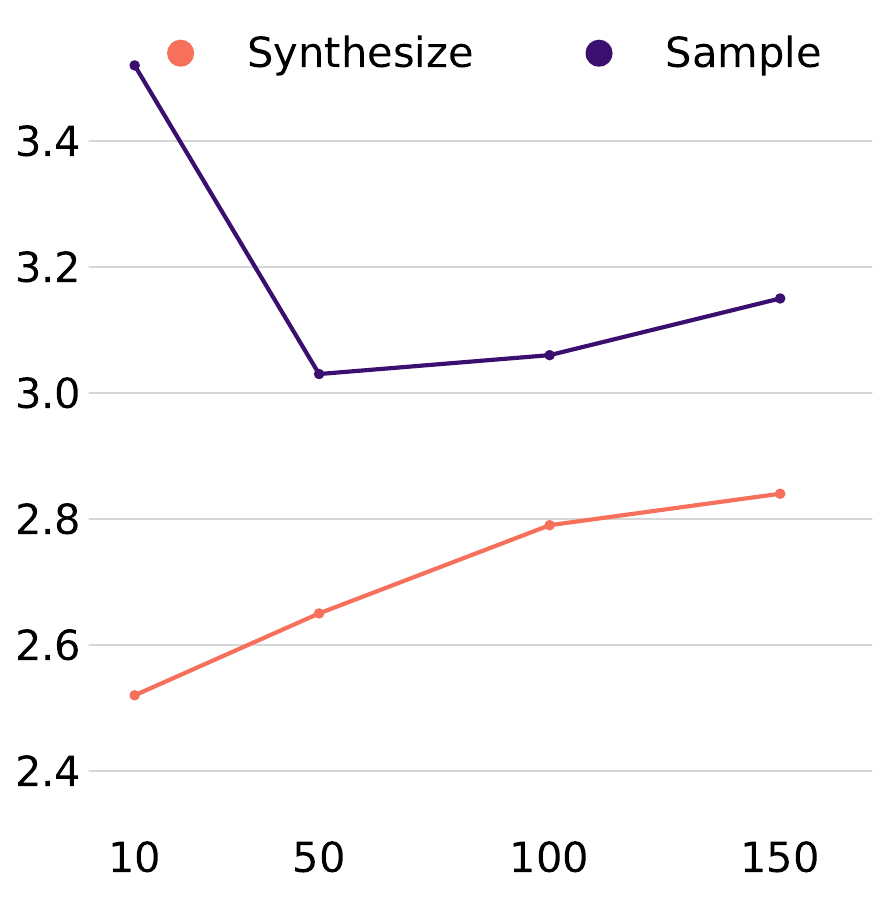}}
\end{tabular}
\caption{Ablation study on number of calibration frames. We compare our noise synthesis method with direct sampling of calibration data at different counts. We show results on MuSCAT-3 for supervised training and MuSCAT-4 on zero-shot inference.}
\label{fig:abaltion_num_frames}
\end{figure*}

\subsubsection{Sensitivity to the Number of Calibration Frames}

We examine how the number of calibration frames affects learning noise characteristics by evaluating denoising performance on the $\topprime{G}$-band, and compare against the direct sampling strategy used in NMOH~\cite{li2025noise}. \Cref{fig:ablation:num:a,fig:ablation:num:b} show a consistent improvement as more calibration frames are available; in the low-data regime, direct sampling can achieve higher performance, whereas physics-based synthesis requires sufficient calibration data to reliably parameterize the statistical model. In contrast, \Cref{fig:ablation:num:c,fig:ablation:num:d} indicate that direct sampling is prone to overfitting to the observed noise realizations, resulting in elevated background noise relative to the physics-based approach. Given the practical scarcity of calibration frames, we use 100 frames throughout all experiments, while fitting more calibration frames can still consistently improve denoising performance.

\section{Conclusion}
In this study, we propose a physics-based CCD noise formation model and a practical synthesis pipeline that generates realistic noisy RAW observations for supervised denoising. Experiments of both in-domain and cross-instrument settings in our novel corrupted real-world multi-band dataset demonstrate that our method consistently outperforms baseline approaches in both photometric and scientific accuracy.

\subsubsection{Limitation}
Strict noise-free ground-truth is unavailable for real astronomical observations, so our evaluation relies on stacked high-SNR bases and calibrated products as practical proxies. In addition, while we include cross-instrument evaluations, all experiments are conducted on two MuSCAT telescopes. Broader validation across other telescopes, sensors, and observing conditions remains an important direction for future studies.

\section*{Acknowledgements}
This work was partially supported by JST Moonshot R\&D Grant Number JPMJPS2011, CREST Grant Number JPMJCR2015 and Basic
Research Grant (Super AI) of Institute for AI and Beyond of the University of Tokyo. In addition, this work was also partially supported by JST SPRING, Grant Number JPMJSP2108.

%
%
\bibliographystyle{splncs04}
\bibliography{main}
\end{document}